\documentclass[showpacs,preprintnumbers,amsmath,amssymb,twocolumn,
prd,a4paper,floatfix,nofootinbib,superscriptaddress,showkeys]{revtex4-1} 
\usepackage{graphicx}                                % Used for includegraphics
\usepackage{amsmath,amsfonts}
\usepackage[caption=false]{subfig}
\graphicspath{{./Figures/}}                          % location for color fig.

\usepackage{amssymb}
\usepackage{amsfonts}
\usepackage{units}
\newcommand{\be}{\begin{equation}}
\newcommand{\ee}{\end{equation}}

\newcommand{\dthree}[1]{\textrm{d}^3#1\,}

\renewcommand{\d}[1]{\textrm{d}#1\,}

\newcommand{\kslash}{k\hspace{-2mm}\slash}
\newcommand{\pslash}{p\hspace{-2mm}\slash}

\newcommand{\vecpslash}{\vec p\hspace{-2mm}\slash}
\newcommand{\veckslash}{\vec k\hspace{-2mm}\slash}
\renewcommand{\vec}[1]{\boldsymbol{#1}}

\newcommand{\tr}[1]{\,\text{tr}\left[#1\right]}

\newcommand{\dless}[1]{\hat{#1}}

\newcommand{\mhat}{\dless{m}}

\newcommand{\phat}{\dless{p}}
\newcommand{\qhat}{\dless{q}}

\newcommand{\SU}[1]{$\textrm{SU}(#1)$}
\renewcommand{\P}{\mathcal{P}}
\newcommand{\intpfour}{\int_{-\pi}^\pi \frac{\d{\phat_4}}{2\pi}}
\renewcommand{\bar}{\overline}

\renewcommand{\Re}{\mathfrak{Re}\,}

\newcommand{\del}{\partial}

\newcommand{\1}{\mathrm{1\hspace{-0.4em}1}}
\newcommand{\A}{\mathcal{A}}
\newcommand{\B}{\mathcal{B}}
\newcommand{\Dcal}{\mathcal{D}}

\renewcommand{\Re}{\mathfrak{Re}\,}
\newcommand{\beq}{\begin{equation}}
\newcommand{\eeq}{\end{equation}}
\newcommand{\bea}{\begin{eqnarray}}
\newcommand{\eea}{\end{eqnarray}}
\newcommand{\beas}{\begin{eqnarray*}}
\newcommand{\eeas}{\end{eqnarray*}}
\newcommand{\eq}{\begin{equation}}
\newcommand{\en}{\end{equation}}
\newcommand{\eqa}{\begin{eqnarray}}
\newcommand{\ena}{\end{eqnarray}}

\newcommand{\Eq}[1]{Eq.~(\ref{#1})}

\begin{document}

\title{Running mass, effective energy and confinement: the lattice quark 
propagator in Coulomb gauge}
%--------------------------------------------------------
\author{G.~Burgio}
\affiliation{Institut f\"ur Theoretische Physik, Auf der Morgenstelle 14, 72076 T\"ubingen, Germany}
\author{M.~Schr\"ock}
\affiliation{Institut f\"ur Physik, FB Theoretische Physik, Universit\"at
Graz, 8010 Graz, Austria}
\author{H.~Reinhardt}
\affiliation{Institut f\"ur Theoretische Physik, Auf der Morgenstelle 14, 72076 T\"ubingen, Germany}
\author{M.~Quandt}
\affiliation{Institut f\"ur Theoretische Physik, Auf der Morgenstelle 14, 72076 T\"ubingen, Germany}

\begin{abstract}
%--------------
We calculate the lattice quark propagator in Coulomb gauge both from 
dynamical and quenched configurations. 
We show that in the continuum limit both the static and full quark 
propagator are multiplicatively renormalizable. From the propagator we extract 
the quark renormalization function $Z(|\vec{p}|)$ and the running mass 
$M(|\vec{p}|)$ and extrapolate the latter to the chiral limit. We find that 
$M(|\vec{p}|)$ practically coincides with the corresponding Landau gauge 
function for small momenta. The computation of $M(|\vec{p}|)$ can be however 
made more efficient in Coulomb gauge; this can lead to a better 
determination of the chiral mass and the quark anomalous dimension.
Moreover from the structure of the full propagator we can read off an 
expression 
for the dispersion relation of quarks, compatible with an IR divergent 
effective 
energy. If confirmed on larger volumes this finding would allow to extend the 
Gribov-Zwanziger confinement mechanism to the fermionic sector of QCD.
\end{abstract}

\maketitle
%-----------------------

\section{Introduction}

Thanks to considerable progress achieved in variational approaches 
\cite{Szczepaniak:2001rg,Feuchter:2004mk,Feuchter:2004gb}, Hamiltonian 
investigations of 
Yang-Mills theories in Coulomb gauge \cite{Christ:1980ku} 
have seen a renewed interest in the last years. In such setup Gau\ss's 
law on the states is explicitely resolved with the help of the gauge
constraint, without any need to construct the physical Hilbert space 
\emph{explicitly} \cite{Burgio:1999tg}. The results obtained \cite{Epple:2006hv}
agree with the Gribov-Zwanziger confinement scenario 
\cite{Gribov:1977wm,Zwanziger:1993dh,Zwanziger:1995cv} and with the existence 
of a confining potential driven by topological excitations 
\cite{Reinhardt:2008ek,Quandt:2010yq,Reinhardt:2011fq}. Lattice 
calculations have confirmed all qualitative and most quantitative features
of the continuum analysis \cite{Burgio:2008jr,Burgio:2008yg,Burgio:2009xp,Burgio:2012bk}. 

Recently, an extension of variational techniques to full QCD has been 
proposed in Ref.~\cite{Pak:2011wu}. In the present paper we wish, via direct 
lattice calculations, to set the benchmark for future checks of such results 
and put forward further predictions. 

A key issue in all lattice studies of pure Yang-Mills theories in Coulomb 
gauge has been the renormalizability of Green's functions. Indeed it has been 
shown in Refs.~\cite{Burgio:2008jr,Burgio:2008yg,Burgio:2009xp,Burgio:2012bk} that for 
{\it static} correlators a non-perturbative renormalization procedure can be 
defined in the lattice Hamiltonian limit $a_t \to 0$. Such limit must be
either taken explicitly \cite{Burgio:2012bk} by going to anisotropic lattices
\cite{Burgio:2003in}, or it can be circumvented whenever a decoupling of the 
temporal and spatial dependence of the correlators is possible 
\cite{Burgio:2008jr,Burgio:2008yg,Burgio:2009xp}. Lacking a clear picture from perturbation 
theory 
\cite{Watson:2007mz,Reinhardt:2008pr,Andrasi:2010dv}, this is the only setup
we are aware of in which multiplicative renormalizability of
Coulomb gauge correlators seems to be guaranteed.

We will show in this paper that, as for the gluon, the lattice Coulomb 
gauge static quark propagator $S(\vec{p}) = \int d p_4 S(p)$ is 
indeed renormalizable. From $S(\vec{p})$ the renormalization function 
$Z(|\vec{p}|)$ 
and the running mass $M(|\vec{p}|)$ can be extracted. The mass function 
$M(|\vec{p}|)$, in particular, encodes the relevant information on
chiral mass and the anomalous dimension; we will find that it basically 
coincides,
for small momenta, with its Landau gauge counterpart. Furthermore, we also 
demonstrate that the {\it full} Coulomb propagator $S(p)$ has a trivial energy 
dependence. Besides making it renormalizable, this allows for a definition 
of a quark effective energy compatible with the confining properties of the 
theory. 

The paper is organized as follows: Sec.~\ref{sec:setup} describes the general
lattice setup used in our investigation; Sec.~\ref{sec:quark} defines
the observables we will study and the requirements for their renormalizability;
in Sec.~\ref{sec:res} we will give our results and discuss their
consequences, while Sec.~\ref{sec:conc} contains our conclusions and
an outlook for future work.

\section{Lattice setup}
\label{sec:setup}

\subsection{Gauge field configurations}
Our calculations have been performed on nine sets of gauge field 
configurations generated by the MILC collaboration \cite{Bernard:2001av, Aubin:2004wf}, 
made available via the Gauge Connection \cite{DiPierro:2011aa} and analyzed 
using mostly the C++ toolkit \texttt{FermiQCD} \cite{DiPierro:2003sz}.
The configurations were produced with the Symanzik-improved L\"uscher--Weisz 
gauge action \cite{Luscher:1984xn}. Seven out of nine sets include two light 
degenerate ($u$, $d$) and one heavier ($s$) quark flavor, while two are in the 
quenched approximation; all dynamical calculations used the Asqtad improved 
action \cite{Orginos:1999cr}. The parameters of all sets are summarized in 
Table~\ref{tab:setup}; for the reported lattice scales 
and quark masses we refer to Refs.~\cite{Bernard:2001av, Aubin:2004wf}. 
The lower part of the table also contains three sets, \emph{(j)-(k)}, of 
quenched configurations on smaller lattices used for cross checks; they were 
produced with the Wilson action at $\beta =$ 5.7, 6.0 and 6.5.
For set ($i$), due to the I/O handling of \texttt{FermiQCD}, our main 
bottleneck, we could only analyze five configurations in reasonable time; 
although averaging over time slices did considerably increase the statistics, 
auto correlations still remained high. This is reflected in the relatively large 
error bars on final quantities for this lattice; we will further comment on this in 
Sec.~\ref{sec:fullstat}.

Throughout the paper $a$ will denote the lattice spacing, $x$ and $p$ 
refer to the lattice 4-coordinate and momentum, respectively, while $\vec{x}$, 
$\vec{p}$ denote the spatial and $x_4$, $p_4$ the temporal position and 
momentum component. Repeated Greek indices indicate summation over all four 
Euclidean components while repeated Latin indices refer to spatial components 
only.

\begin{table}[htb]
	\begin{tabular}{l l l c c c}\hline\hline
	 & $L^3\times T$ & $a\left[\mathrm{fm}\right]$ & $am$ & $m\left[\mathrm{MeV}\right]$ & \# config. \\\hline
($a$) & $20^3\times64$ & 0.121 & 0.010, 0.050 & 15.7, 78.9 & 202 \\
($b$) & $20^3\times64$ & 0.121 & 0.020, 0.050 & 31.5, 78.9 & 50 \\
($c$) & $20^3\times64$ & 0.120 & 0.030, 0.050 & 47.3, 78.9 & 25 \\
($d$) & $20^3\times64$ & 0.119 & 0.040, 0.050 & 63.1, 78.9 & 25 \\
($e$) & $20^3\times64$ & 0.121 & --           & --          & 66\\\hline
($f$) & $28^3\times96$ & 0.086 & 0.0062, 0.031 & 14.0, 67.8 & 50 \\
($g$) & $28^3\times96$ & 0.086 & 0.0124, 0.031 & 27.1, 67.8 & 50 \\
($h$) & $28^3\times96$ & 0.086 & --            & --          & 77 \\\hline
($i$) & $64^3\times144$ & 0.060 & 0.0036, 0.0108 & 11.8, 35.3 & 5 \\\hline
($j$) & $16^3\times32$ & 0.170 & --             & --         & 94 \\
($k$) & $16^3\times32$ & 0.093 & --             & --          & 108 \\
($l$) & $16^3\times32$ & 0.045 & --            & --         & 131 \\\hline\hline
	\end{tabular}
	\caption{The gauge field configurations used in this study. 
Sets ($a$)--($d$) and ($f$)--($g$) include two degenerate 
($u$, $d$) quarks and one heavier ($s$) flavour 
\cite{Bernard:2001av,Aubin:2004wf}.}
\label{tab:setup}
\end{table}

\subsection{Gauge fixing}
\label{sec:gf}

The continuum Coulomb gauge condition,
\be
\vec{\nabla}\cdot \vec{A} = 0
\ee
can be realized on the lattice by maximizing the gauge functional
\be
F_g[U] = \Re\sum_{ i, x} \mathrm{tr}\big[ U^g_ i(x) + U^g_ i(x-\hat i)^\dagger\big]
\label{eq:F}
\ee
with respect to gauge transformations $g(x)\in\mathrm{SU}(3)$, where
\be
 U^g_\mu(x) \equiv g(x) U_\mu(x) g(x+\hat\mu)^\dagger.
\ee
The gauge links maximizing Eq.~(\ref{eq:F}) will satisfy
the discretized Coulomb gauge condition
\be
\Delta^g(x)\equiv \sum_i\left(A_i^g(x)-A_i^g(x-\hat i) \right) =0\,,
\ee
where we define the gauge fields as:
\be
A_i(x)\equiv \left[\frac{U_i(x)-U_i(x)^\dagger}{2iag_0}\right]_{\textrm{traceless}}.
\ee
This corresponds to Hermitian generators $T^a = \lambda^a / 2$ for the 
Lie algebra of $SU(3)$, where $\lambda^a$ are the Gell-Mann matrices.

To maximize $F_g[U]$ we employ the standard over-relaxation algorithm 
\cite{Mandula:1990vs}, which in the case at hand can be applied 
to each time-slice independently. The inversion of the Dirac operator still 
needs however to be performed on the whole lattice and is 
computationally quite expensive. This forces us to analyze less configurations
than those used, say, to calculate the Coulomb operator \cite{Burgio:2012bk},
making statistics lower and raising the ratio of 
statistical noise to Gribov noise; we have therefore not 
seen, at least for this first study, any need to adopt the improved techniques 
first developed for Landau gauge in 
Refs.~\cite{Bogolubsky:2005wf,Bogolubsky:2007bw} and adapted to
Coulomb gauge in Refs.~\cite{Burgio:2008jr,Burgio:2008yg,Burgio:2009xp,Burgio:2012bk}.
Due to the presence of fermions the part of such techniques based on 
center transformations and thus topologically non-trivial $\mathbb{Z}_N$ 
sectors 
\cite{Bogolubsky:2005wf,Bogolubsky:2007bw} could have not been applied
anyway. 

A measure of the quality of the gauge fixing for each fixed time-slice 
is the average $L_2$-norm of the gauge fixing violation $\Delta^g \neq 0$ 
\cite{Giusti:2001xf}
\be
\theta(x_4) \equiv \frac{1}{L^3N_c}\sum_{\vec x}\tr{\Delta^g(\vec{x},x_4)
\Delta^g(\vec{x},x_4)^\dagger}\,,
\ee
where the sum runs over all spatial sites $\vec x$ and $L^3$ is the 
number of lattice sites in one time-slice.
We have chosen to stop the over-relaxation algorithm whenever on each time-slice
$\sqrt{\theta}$, the default output of the \texttt{FermiQCD} code, reached machine 
single-precision, $\sqrt{\theta} \lesssim 5\, 10^{-7}$. 
Notice that in the literature (see e.g. Ref.~\cite{Parappilly:2005ei}) 
the value of $\theta$ itself, rather than of $\sqrt{\theta}$, is often reported 
for the quality of the gauge fixing; in our case this corresponds to 
${\theta} \lesssim 2.5\, 10^{-13}$. A possible improvement we are considering to implement
in the future is to introduce a local rather than a 
global observable in triggering the stopping criterion, see e.g. 
Ref.~\cite{Bogolubsky:2005wf,Bogolubsky:2007bw}, resulting in a better gauge 
fixing.

As usual in Coulomb gauge maximizing Eq.~(\ref{eq:F}) leaves the temporal links 
$U_4(x)$ unfixed, i.e.~we still have a gauge freedom with respect to 
space independent gauge transformations $g(x_4)\in$ \SU{3}, which leave 
$F_g[U]$ unaffected. One possible choice to fix the residual gauge in the 
continuum is to require
\be
\del_4 \int \dthree{x} A_4(x) =0\,.
\label{eq:tgauge}
\ee
Throughout this paper we will use the lattice version of Eq.~(\ref{eq:tgauge})
proposed in Ref.~\cite{Burgio:2008jr,Burgio:2008yg},
which we here call integrated Polyakov gauge (IPG) and which we have adapted 
to the colour group $SU(N)$. For sufficiently fine temporal lattice spacing 
the IPG might be considered a good approximation, on average, 
to the Weyl gauge $A_4 = 0$ used for the Hamiltonian approach in the continuum.
For further details see Appendix~\ref{app:tgauge}.

\section{The quark propagator}
\label{sec:quark}

\subsection{Quark propagator in Coulomb gauge}
At tree-level the inverse continuum quark propagator in Euclidean space 
reads:
\be
	S^{(0)}(p)^{-1} = i \vecpslash  + i \pslash_4 +  m\,,
\ee
where we have used Feynman's slash notation, 
$\vecpslash \equiv \sum_i \gamma_i p_i$ and $\pslash_4\equiv \gamma_4 p_4$;
$m$ denotes the bare quark mass.  
We have explicitly separated the spatial momenta $p_i$ from the temporal 
component $p_4$ (the energy) to make contact with the 
non-manifestly Euclidean invariant interacting Coulomb gauge propagator 
$S^{-1}(p)$. 
The latter could, in principle, have a more complex Dirac structure than in 
covariant gauges since three-dimensional (spatial) covariance allows 
for more possible contractions of the momentum components with elements of the 
Euclidean Clifford algebra. More precisely, the spatial momentum $\vec{p}$ 
can be contracted with both the spatial Dirac matrices $\gamma_i$ and the 
proper vector component $\sigma_{i4}$ of the Euclidean tensor
$\sigma_{\mu\nu} = \nicefrac{i}{2}[\gamma_\mu, \gamma_\nu]$. Thus, 
$S^{-1}(p)$ can be decomposed into four pieces:
\bea
\label{dress}
S^{-1}(p) &=& i \vecpslash A_s(|\vec{p}|,p_4)  + 
i \pslash_4 A_t(|\vec{p}|,p_4) \nonumber\\
&+& i \sum_j p_j \sigma_{j4} A_d(|\vec{p}|,p_4) +  B_m(|\vec{p}|,p_4)
\label{eq:quark4}
\eea
with scalar functions $A_s(|\vec{p}|,p_4)$, $A_t(|\vec{p}|,p_4)$, 
$A_d(|\vec{p}|,p_4)$ and $ B_m(|\vec{p}|,p_4)$,  
to which we will refer
as the spatial, temporal, mixed and massive component, respectively.

For our calculations of the valence quark propagator we have used Asqtad 
improved \cite{Orginos:1999cr} staggered fermions \cite{Kogut:1974ag}, 
concentrating on the dynamical point for the unquenched configurations. 
In Landau gauge this choice
\cite{Bowman:2002bm,Parappilly:2005ei,Bowman:2001xh,Bowman:2002kn,%
Bowman:2005vx,Furui:2005mp}
showed not only good agreement with the conceptually cleaner but very 
expensive overlap fermions 
\cite{Bonnet:2002ih,Zhang:2003faa,Zhang:2004gv,Kamleh:2004aw,Kamleh:2007ud,%
Bowman:2005zi, Bowman:2004xi}
but also suffered less from short-distance cutoff effects as compared to clover 
(improved Wilson) \cite{Skullerud:2000un, Skullerud:2001aw}
or chirally improved (CI) fermions \cite{Schrock:2011hq}.

Only for the finest lattice (set $(i)$) we were forced to use standard 
Kogut--Susskind fermions \cite{Kogut:1974ag} due to memory limitations. 
For this set we have calculated the propagator with a mass equal
to the two light degenerate dynamical quarks (\unit[11.8]{MeV}) plus four 
partially quenched masses up to \unit[142.1]{MeV}.
On the quenched configuration ($h$) we have used a mass of \unit[14.0]{MeV}, 
i.e.~the same as on configuration ($f$), in order to study the effects of 
dynamical quarks.

Notice that for standard staggered fermions $S(p)$ will actually be a function
of $k_\mu\equiv\sin p_\mu$, while the equivalent
expression for the Asqtad improved quarks can be found in 
Appendix~\ref{staggDetails}. To keep notations simple we will however write 
the structure functions $A_s$, $A_t$, $A_d$ and $B_m$ as functions of the the 
discrete momenta $p_\mu$ taking values in the first Brillouin zone.

\subsection{Dispersion relation and renormalizability}
\label{sec:ren}
From the configurations at our disposal, once any suitable gauge
has been fixed, the inversion of the Dirac operator directly provides us 
with the regularized propagator $S_{\mathrm{reg}}(a;p)$,
which depends on the lattice spacing $a$. Assuming multiplicative 
renormalizability, such regularized propagator should be related to the 
renormalized one $S_\zeta(p)$ via the quark wavefunction renormalization 
constant $Z_2$, which will depend on $a$ and the renormalization point 
$\zeta$,
\be
S_{\mathrm{reg}}(a;p) = Z_2(\zeta;a)S_\zeta(p)\,.
\label{eq:Sren}
\ee
In Coulomb gauge the static propagator can be then extracted from $S_\zeta(p)$ 
by integrating it over $p_4$.
For bosonic fields such static propagator agrees,
up to a constant, with the inverse of the boson's dispersion relation 
$\omega_B(|\vec{p}|)$. This connection is essential in showing e.g. that 
the Gribov-Zwanziger confinement scenario for the gluon is indeed 
realized within pure Yang-Mills theories \cite{Burgio:2008jr,Burgio:2008yg,Burgio:2009xp}. 

In the fermionic case things are a bit more complicated. In Landau 
gauge, up to well understood discretization effects, the renormalized 
propagator $S_\zeta(p)$ was shown \cite{Bowman:2002bm,Parappilly:2005ei,%
Bowman:2001xh,Bowman:2002kn,Bowman:2005vx,Furui:2005mp,Schrock:2011hq}
to have indeed the expected form
\be
S_\zeta(p) = \frac{Z_\zeta(p^2)}{i \pslash + M(p^2)}\,,
\label{eq:dec}
\ee
where the only dependence on the renormalization scale $\zeta$ is through
the numerator $Z_\zeta(p^2)$, while the mass function $M(p^2)$ in the 
denominator does not depend on the cutoff $a$ or the renormalization scale 
$\zeta$;
it is thus a renormalization group invariant. 

What is the physical
meaning of such propagator and where can the dispersion relation
be read from? At least for the non interacting case, 
where $M\equiv m$ and $Z$ is constant, we know that the 
inverse propagator Eq.~(\ref{eq:dec})
is proportional, up to a Wick rotation, to the projectors on the positive and 
negative energy states \cite{Itzykson:1980rh}:
\be
\Lambda_\pm (p) \propto {\pm \pslash +m};
\label{eq:proj}
\ee 
the free particle dispersion relation can now be obtained by integrating the 
\emph{square} of such Euclidean propagator in the energy $p_4$, 
\be
\omega^{-1}_{F}(|\vec{p}|) = 2 \int \frac{d p_4}{2 \pi} 
\frac{1}{p_4^2 + \vec{p}^2 + m^2} = \frac{1}{\sqrt{\vec{p}^2 + m^2}}.
\label{eq:sq_free}
\ee
On the other hand, Eq.~(\ref{eq:dec}) in the interacting case can be considered 
the propagator of a single quasi-particle, where $M$ depends on the 
4-momentum $p$. In this interpretation a projector as in Eq.~(\ref{eq:proj}) 
can still be defined by simply substituting $m \to M(p^2)$. To extract the 
effective energy of the quasi-particle, however, one needs to 
model the functional form of $M$ and $Z$ in order to perform the integral as in 
Eq.~(\ref{eq:sq_free}), since a numerical summation over $p_4$ would be
plagued by cut-off effects \cite{Burgio:2008jr,Burgio:2008yg}.

In Coulomb gauge, we can study both the full energy-dependent propagator 
as in Eq.~(\ref{eq:quark4}) and the \emph{static} propagator 
$S(\vec{p}) = \int d p_4 S(p)$, taking the form (see Sec.~\ref{sec:fullstat}):
\be
S_\zeta(\vec{p}) = \frac{Z_\zeta(|\vec{p}|)}{i \vecpslash + M(|\vec{p}|)}\,.
\label{eq:decst}
\ee 
What is the meaning of these quantities? Consider the free Dirac equation:
\be
i \frac{\partial}{\partial t} \psi = p_4 \psi = {h(\vec{p})} \psi = 
\gamma_4 \left(\vecpslash + m\right)\psi\,,
\label{eq:DirHam}
\ee
where the one-particle Hamiltonian $h(\vec{p})$ coincides, up to a 
gamma matrix and a Wick rotation, with the free inverse static 
propagator (i.e.~$M\equiv m$ and $Z$ constant). We can again define 
the projectors $\tilde{\Lambda}_\pm$ on the positive/negative free energy 
states and find upon Wick rotation:
\bea
\tilde{\Lambda}_+(\vec{p}) - \tilde{\Lambda}_-(\vec{p}) &\propto& {h(\vec{p})}
\propto S^{-1}(\vec{p})\nonumber\\
\tilde{\Lambda}_+(\vec{p}) + \tilde{\Lambda}_-(\vec{p}) &=&1\,,
\eea
while the dispersion relation is given by the eigenvalues of $h(\vec{p})$.
If we now turn on the interactions and appeal again to the quasi-particle
description
we will have in general:
\be
h(\vec{p}) = \frac{1}{\rho(|\vec{p}|)}\gamma_4 \left(\vecpslash 
+ M(|\vec{p}|)\right)
\ee
and the (Wick rotated) static Coulomb propagator will still define the energy 
projectors. 
Moreover, if we could show for the full Euclidean Coulomb propagator a relation
of the form:
\be
S_\zeta(p) = \frac{Z_\zeta(|\vec{p}|)}{i \vecpslash + i \pslash_4
\alpha(|\vec{p}|)+ M(|\vec{p}|)}
\label{eq:renns}
\ee
we could, up to a constant, directly read from the integration of $S^2$ in 
$p_4$ the dispersion relation i.e. the effective energy of the quasi-particle:
\bea
\omega^{-1}_F(|\vec{p}|) &=& 2 Z_\zeta^2(|\vec{p}|) \int \frac{d p_4}{2 \pi} 
\frac{1}{\alpha^2(|\vec{p}|) p_4^2 + \vec{p}^2 + M^2(|\vec{p}|)}\nonumber\\ 
&=& 
\frac{Z_\zeta^2(|\vec{p}|)}{\alpha(|\vec{p}|)\sqrt{\vec{p}^2 + 
M^2(|\vec{p}|)}}\,.
\label{eq:effene}
\eea

To check the above conjectures we first need to extract the four 
dressing functions in \Eq{dress} from the lattice calculation. This can be done 
by extending the techniques developed for Wilson and staggered fermions in 
Landau 
gauge in Refs.~\cite{Bowman:2002bm, Bowman:2005vx}. 
For sake of readability all technical details have been deferred to Appendix 
\ref{staggDetails}, where the staggered case is explicitely 
discussed. The adaptation of the method to Wilson and Asqtad fermions 
is straightforward.

\begin{figure*}[htb]
\centering
\subfloat[Configs. ($a$)]{\label{2064_A_d_c}\includegraphics[width=0.8\columnwidth]{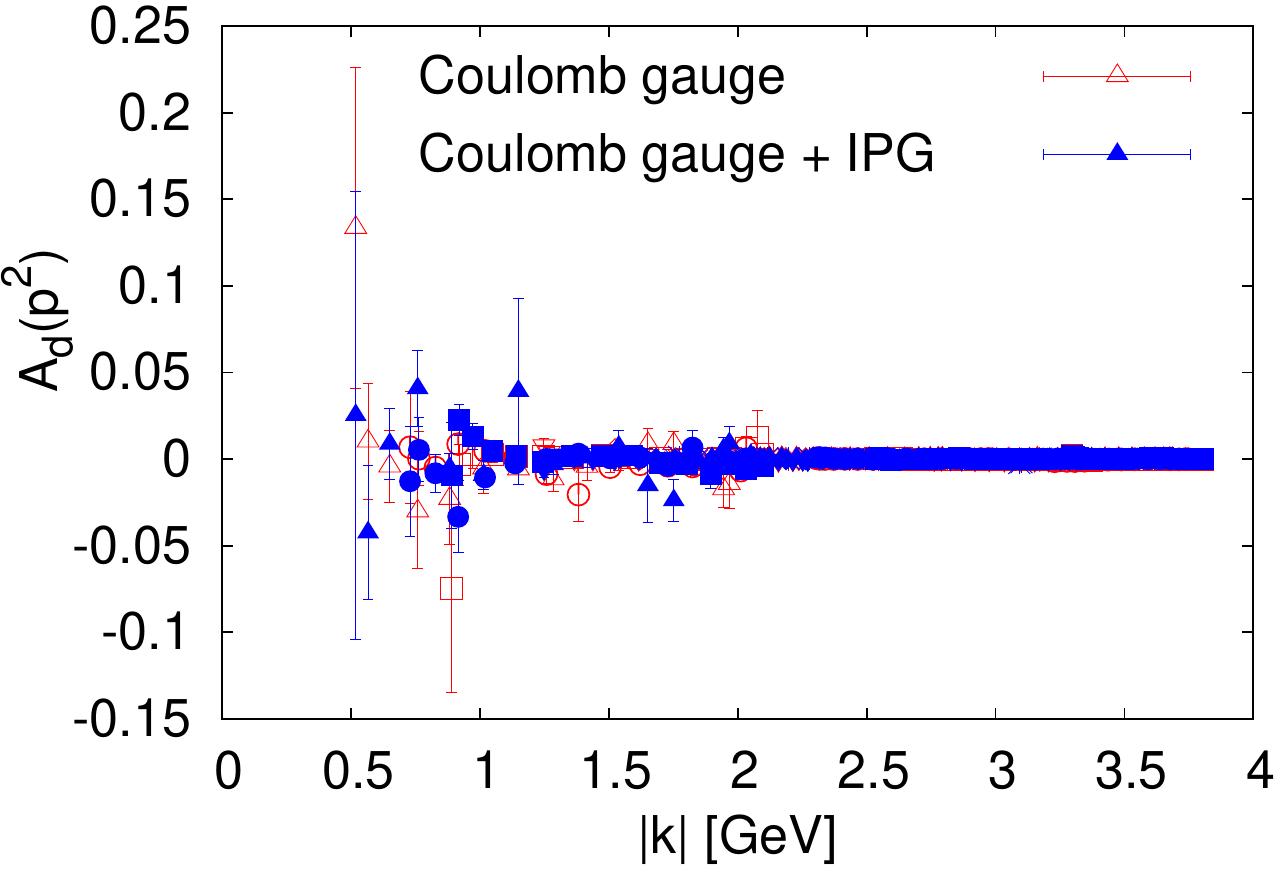}} \hspace*{1cm}
\subfloat[Configs.~($a$), $p_4$ averaged]{\label{2064_A_d_d}\includegraphics[width=0.8\columnwidth]{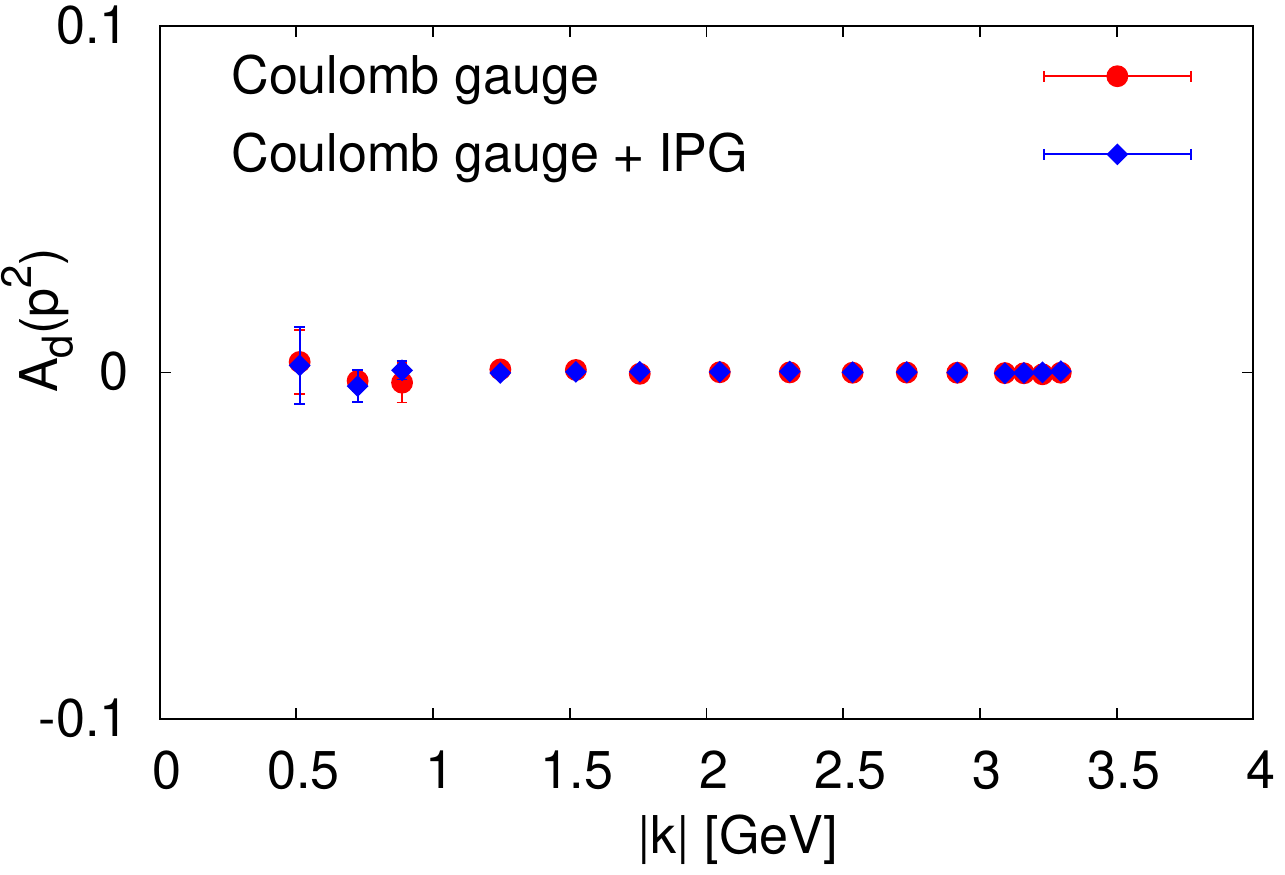}}\\
\subfloat[Configs. ($b$), $p_4$ averaged]{\label{2064_A_d_a}\includegraphics[width=0.9\columnwidth,height=5truecm]{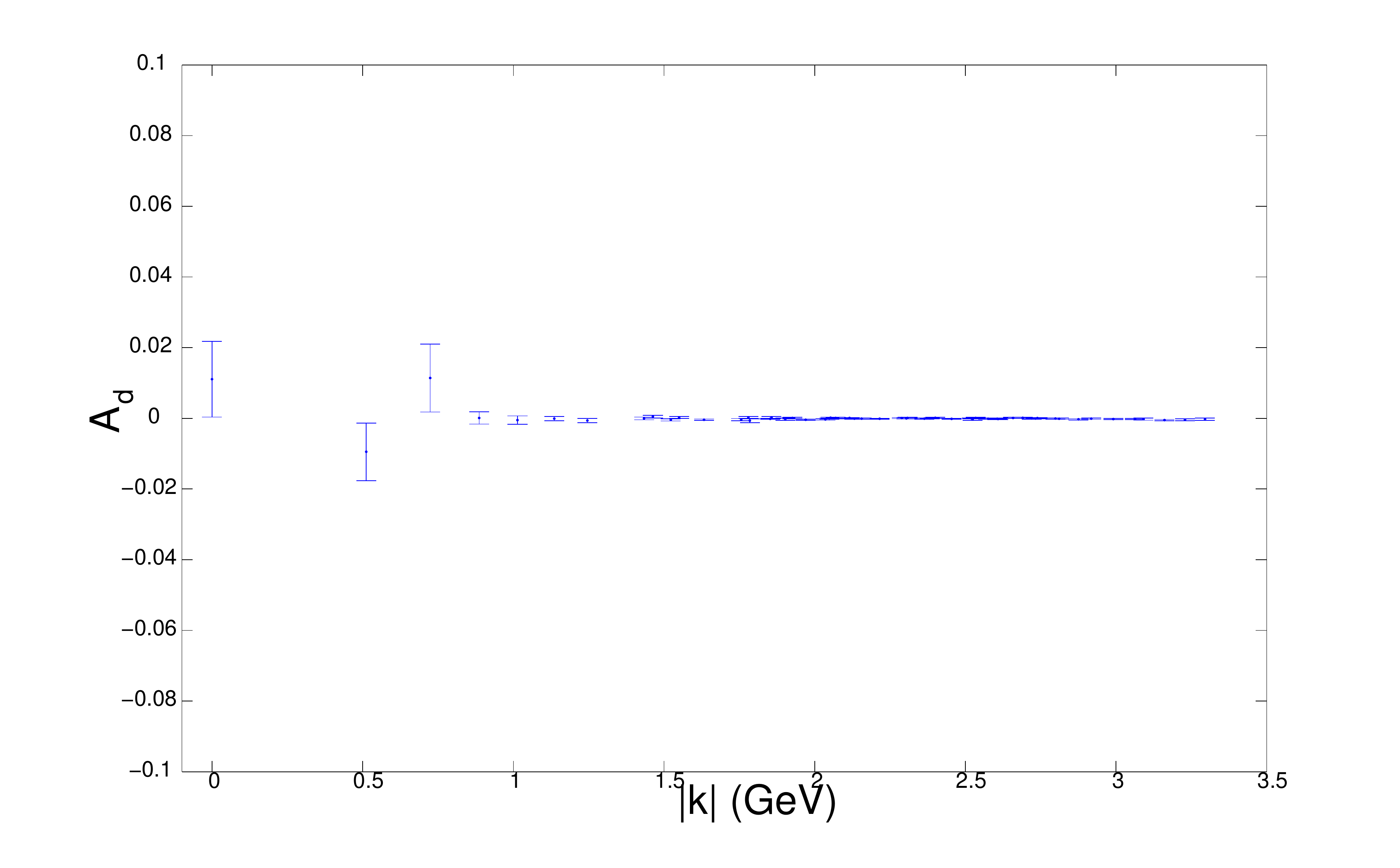}} 
\subfloat[Configs. ($k$), $p_4$ averaged]{\label{2064_A_d_b}\includegraphics[width=0.9\columnwidth,height=5truecm]{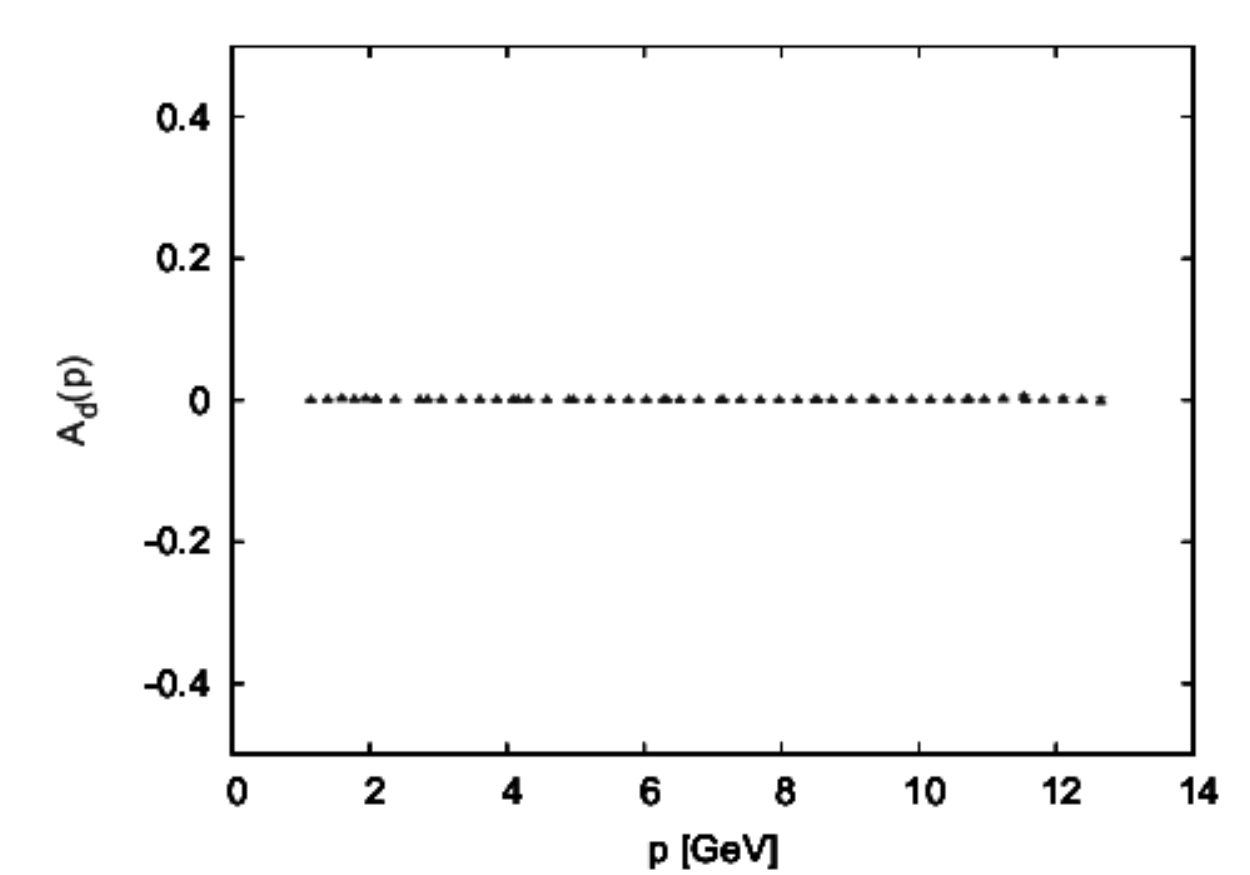}}
\caption{Dressing function $A_d$. In Subfigs.~\ref{2064_A_d_c},~\ref{2064_A_d_d}
the effect of residual gauge fixing (IPG) is shown.}
\label{2064_A_d}
\end{figure*}

\subsection{Data cuts and averaging}
Throughout this work we average over the cubic (spatial) symmetries of the 
lattice and the parity symmetry of the propagator $S(p)=S(-p)$.
To minimize discretization artifacts caused by the breaking of rotational 
invariance we perform a cylinder cut on the spatial momenta
\cite{Leinweber:1998uu,Voigt:2007wd}.

\section{Results}
\label{sec:res}

\subsection{Dirac structure}

Fig.~\ref{2064_A_d} shows the structure function $A_d$ for some of the data set
in Tab.~\ref{tab:setup}, plus one of the quenched Wilson simulation, set $(k)$ 
(Fig.~\ref{2064_A_d_b}), used as a cross check. In all cases $A_d \equiv 0$, 
which 
extends the one loop result of Ref.~\cite{Popovici:2008ty} to the
non perturbative regime. We will thus ignore the mixed component 
in the rest of this paper and express the lattice propagator as:
\be
S^{-1}(p) =  i\veckslash a A_s(|\vec{p}|,p_4)+ 
i \kslash_4 a A_t(|\vec{p}|,p_4)+ B_m(|\vec{p}|,p_4)\,,
\label{eq:prop_nm}
\ee
where we denote with $k_\mu$ the dimensionless lattice momenta (see 
Appendix~\ref{staggDetails}) while the spacing $a$ in the first two terms
renders the dressing functions $A_s$ and $A_t$ dimensionless. 

\subsection{Energy dependence}
\label{sec:end_dip}

\begin{figure*}[htb]
\centering
\subfloat[$d_m$ from ($a$)]{\label{2064_d_a}\includegraphics[width=0.80\columnwidth,height=5truecm]{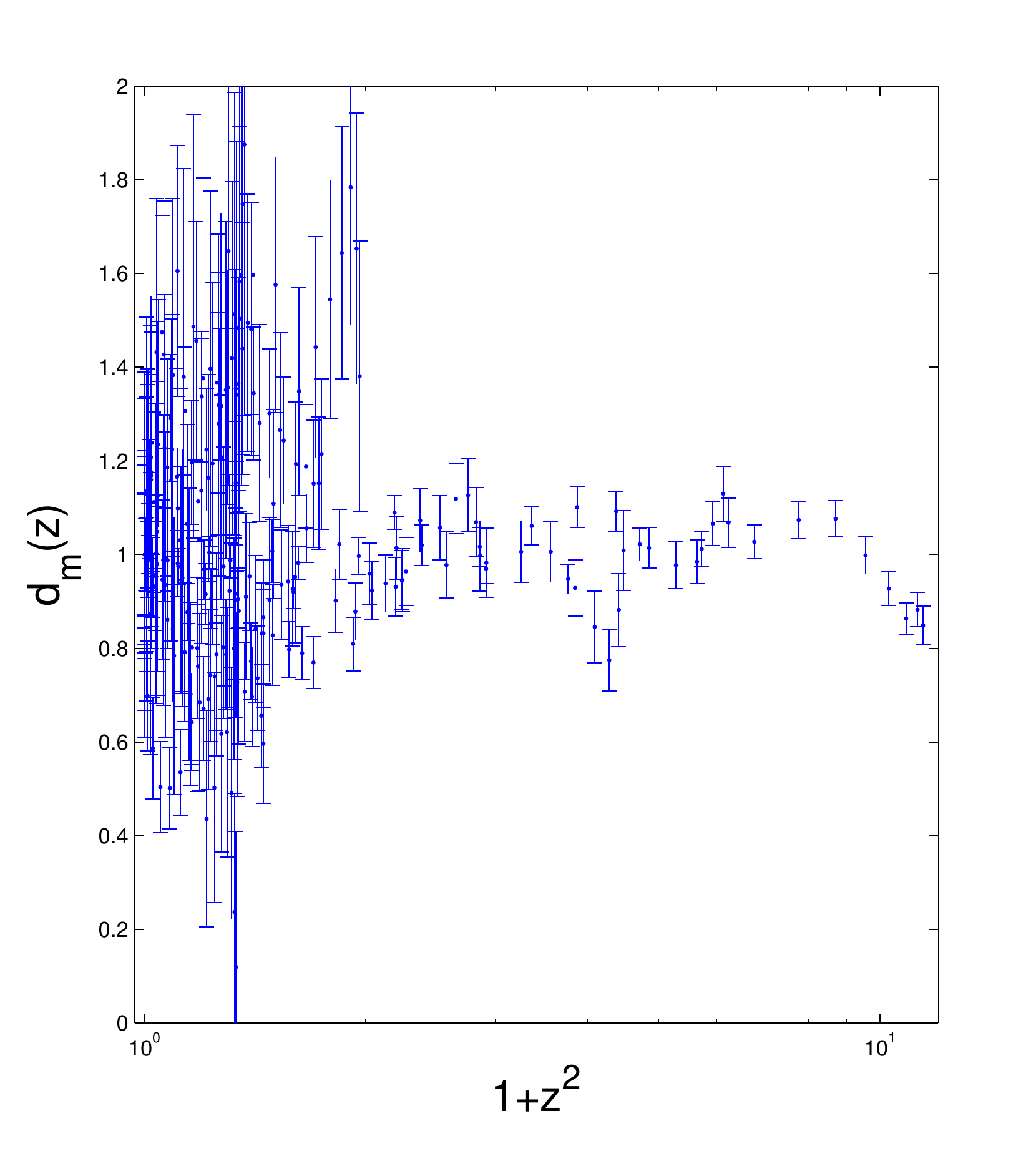}}
\subfloat[$d_t$ from ($a$)]{\label{2064_d_b}\includegraphics[width=0.80\columnwidth,height=5truecm]{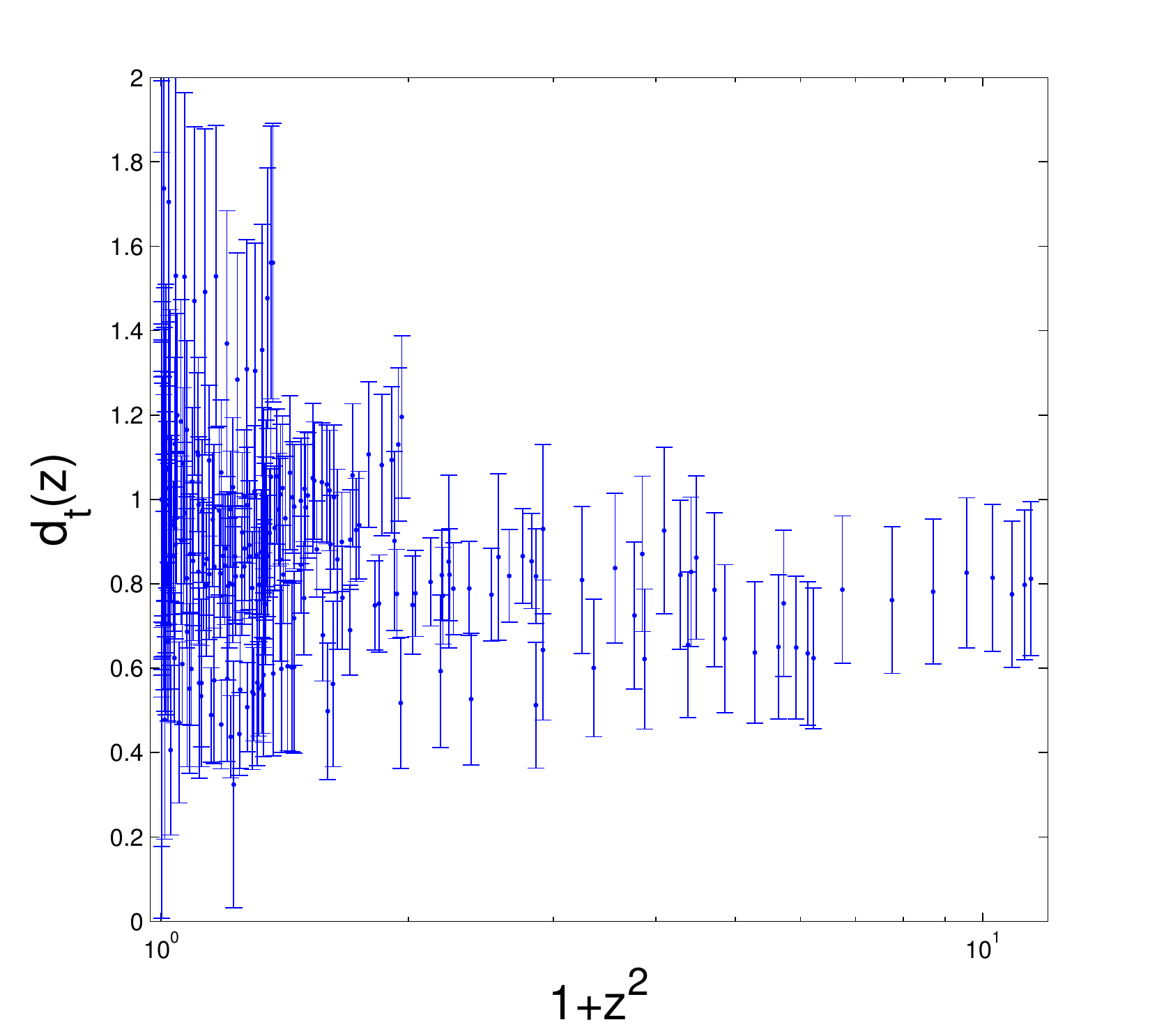}}\\
\subfloat[$d_s$ from ($a$)]{\label{2064_d_c}\includegraphics[width=0.80\columnwidth,height=5truecm]{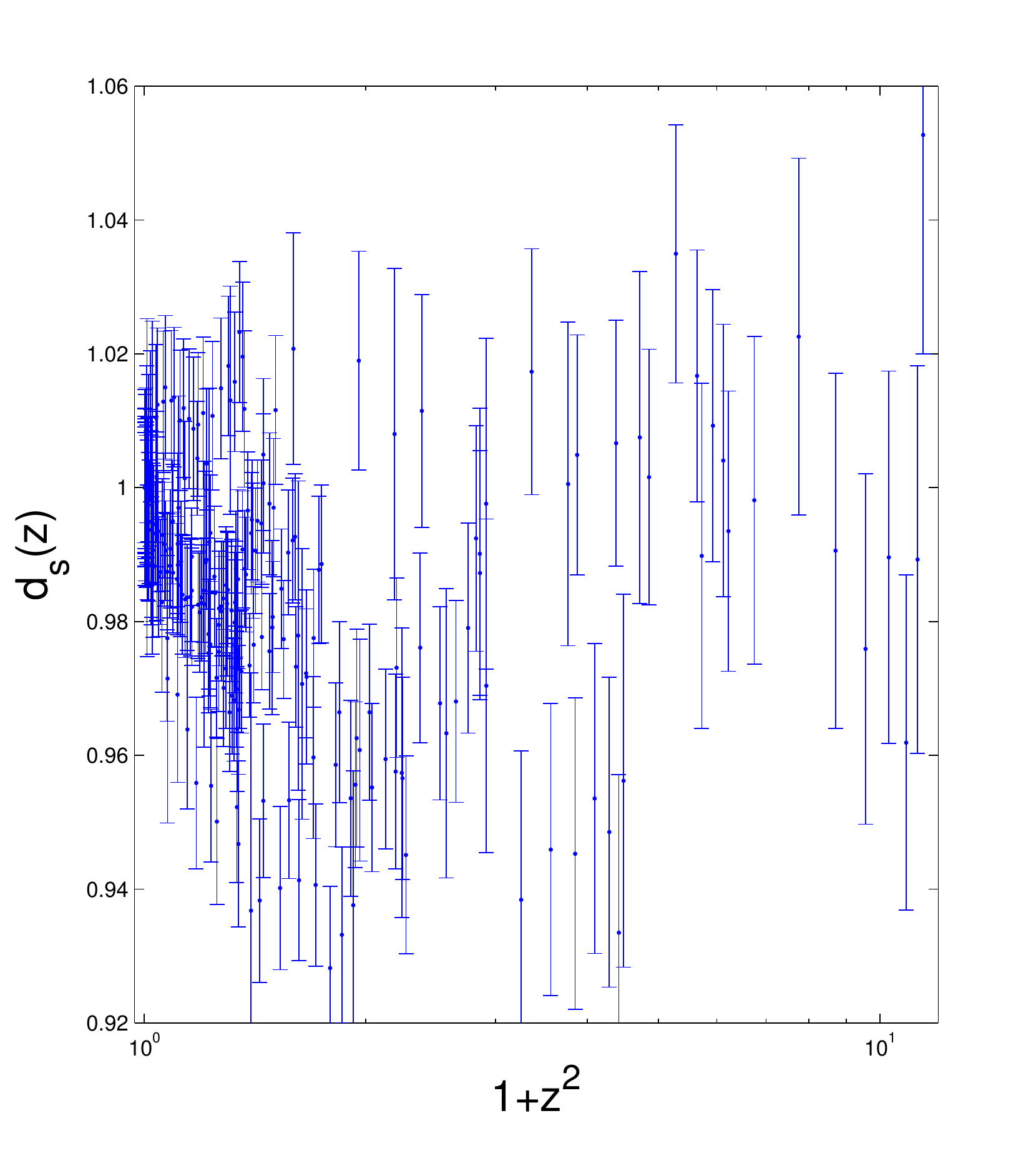}}
\subfloat[$d_s$ from ($i$)]{\label{2064_d_d}\includegraphics[width=0.80\columnwidth,height=5truecm]{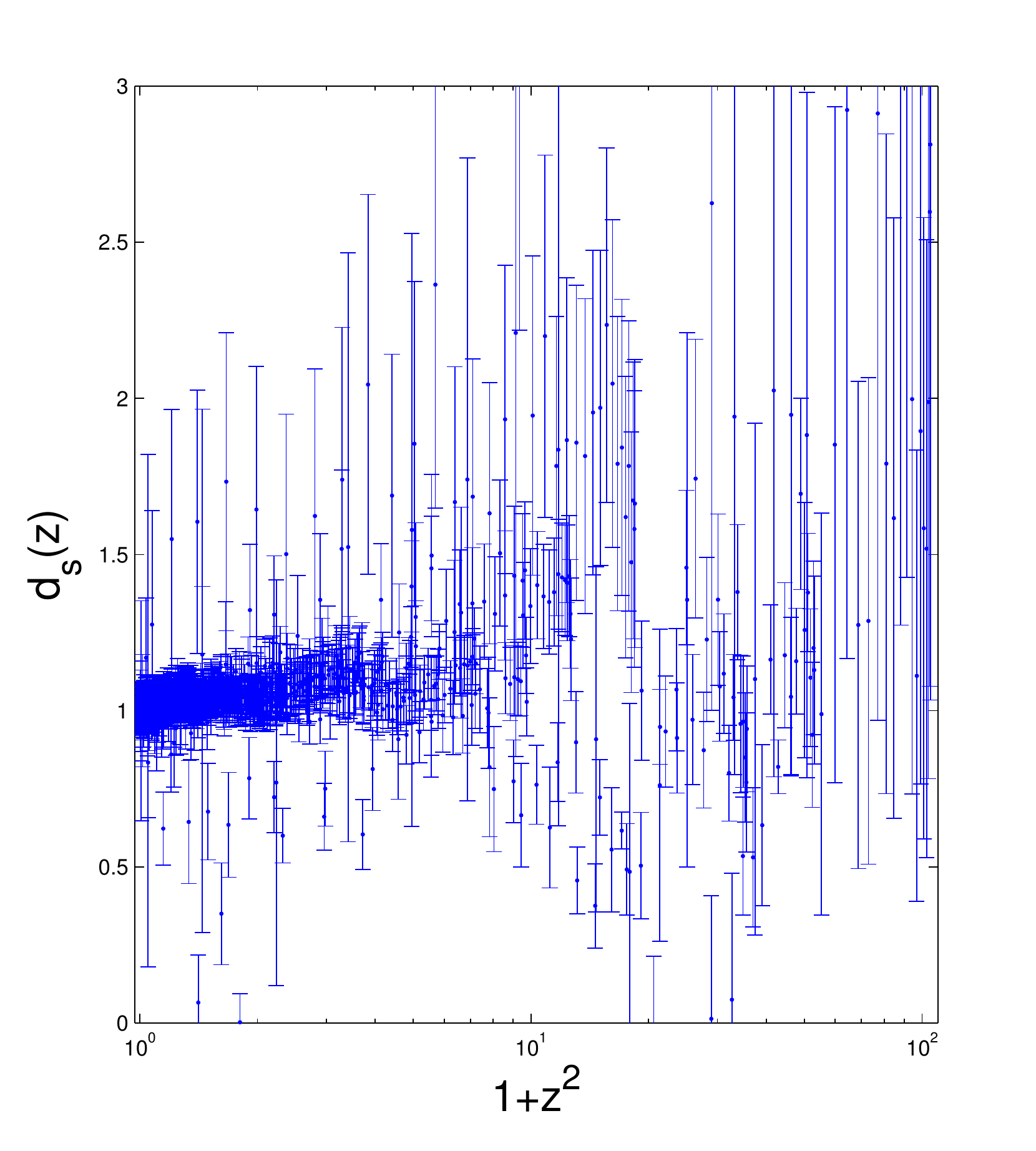}}
\caption{Energy dependence of dressing functions as defined in Eq.~(\ref{eq:d}).}
\label{2064_d}
\end{figure*}

In Refs.~\cite{Burgio:2008jr,Burgio:2008yg,Burgio:2009xp,Burgio:2012bk} it was shown that 
the static propagators of pure lattice gauge theory in Coulomb gauge are 
subject to cutoff effects in the temporal lattice spacing $a_t$, which lead
to  violations of multiplicative renormalizability. These effects can be both 
direct, i.e.~caused by an explicit energy dependence of the correlator at hand 
\cite{Burgio:2008jr,Burgio:2008yg,Burgio:2009xp}, and indirect, i.e.~caused by $\mathcal{O}(a_t)$ 
corrections to the spectrum of the theory \cite{Burgio:2012bk}. In principle 
we cannot exclude the latter effect for the quark propagator. 
However, the indirect energy dependence is usually much smaller than the direct
effect, and a higher statistical precision as that in the present work, 
combined 
with simulations on anisotropic lattices \cite{Burgio:2003in}, would be 
required to resolve it. As for the direct energy dependence of the quark 
dressing functions, a quantitative measure is given by:
\bea
d_m(z) &=& \frac{B_m(|\vec{p}|,p_4)}{B_m(|\vec{p}|,p_4^{min})}\nonumber\\
d_t(z) &=& \frac{A_t(|\vec{p}|,p_4)}{A_t(|\vec{p}|,p_4^{min})}\nonumber\\
d_s(z) &=& \frac{A_s(|\vec{p}|,p_4)}{A_s(|\vec{p}|,p_4^{min})}\,,
\label{eq:d}
\eea
which on dimensional grounds should only be functions of 
$z = \frac{p_4}{|\vec{p}|}$. In Fig.~\ref{2064_d} we show as
an example the functions 
of Eq.~(\ref{eq:d}) for configurations sets ($a$) and ($i$), plotted as a
function of $1+z^2$ to better compare with Fig.~1 of Ref.~\cite{Burgio:2008jr}.

We can conclude that, at least within our numerical precision, the functions 
$A_s$, $A_t$ and $ B_m$ are independent of the energy $p_4$. 
Contrary to the gluon propagator \cite{Burgio:2008jr,Burgio:2008yg,Burgio:2009xp},  
explicit violations of renormalizability due to the temporal cutoff 
$a_t$ can therefore be ruled out.
If we do keep $p_4$ as an argument it is only to distinguish the  
$p_4$-un-averaged structure functions obtained after the temporal gauge has 
been  fixed from the $p_4$-averaged ones which describe the static 
functions obtained for a fixed time slice.

\subsection{Full vs. static propagator and their renormalization}
\label{sec:fullstat}

Due to the energy independence of the dressing functions Eq.~(\ref{eq:prop_nm})
we can average them over $p_4$ to minimize statistical
fluctuations. The full propagator thus reads
\be
S^{-1}(p) =  i\veckslash a A_s(|\vec{p}|)+ 
i \kslash_4 a A_t(|\vec{p}|)+ B_m(|\vec{p}|)\,,
\label{eq:prop_nm_ne}
\ee
while its static counterpart is, up to a constant proportional to the 
time extent of the lattice (see Appendix~\ref{A-ZnM}):
\be
S^{-1}(\vec p) = i\veckslash a A_s(|\vec{p}|)
+ B_m(|\vec{p}|)\,.  
\label{eq:intprop}
\ee
Alternatively, the static propagator can be
taken at a fixed time slice, which is equivalent to averaging
over all time slices without residual gauge fixing. As shown in 
Fig.~\ref{2064_al_b}, $A_t$ identically vanishes in this case 
and we again obtain Eq.~(\ref{eq:intprop}). Such fixed time definition would
of course make the static quark propagator much easier to calculate on lattices 
with large temporal extension, since for $L^3 \times T$ the inversion of the 
Dirac  operator can be restricted to the $L^3$ spatial sub-lattice. Even if 
the procedure is repeated $T$ times to improve the statistics one will still 
cut both CPU time and memory requirements and avoid possible I/O inefficiencies. We
are planning to implement such improvement in all future analysis; besides
enabling us to increase the number of configurations, this should
also allow us to invest more computer time in the quality of the gauge fixing, see 
discussion in Sec.~\ref{sec:gf}. Of course, this does not apply to the full
propagator of Eq.~(\ref{eq:prop_nm_ne}), which will still need the inversion
of the whole Dirac operator to resolve $A_t$.

Let us now define the renormalized propagators. While from 
Eq.~(\ref{eq:intprop}) a static 
propagator as in Eq.~(\ref{eq:decst}) immediately follows, an explicit $p_4$ 
dependence of the dressing functions would have made it very difficult 
to cast $S(p)$ to the form Eq.~(\ref{eq:renns}), with the scale 
dependence confined to the proportionality factor. Due to the trivial 
$p_4$-dependence of Eq.~(\ref{eq:prop_nm_ne}), however, the full (non-static)
quark propagator does indeed take the lattice equivalent form of 
Eq.~(\ref{eq:renns}):
\be
S_\zeta(p) = \frac{Z_\zeta(|\vec{p}|)}{i a \veckslash + i a \kslash_4
\alpha(|\vec{p}|)+ M(|\vec{p}|)}\,.
\label{eq:fullprop}
\ee
The
renormalization function $Z_\zeta(|\vec p|)$, the mass function 
$M(|\vec p|)$ and the ``energy form factor''
$\alpha(|\vec p|)$ are derived
in Appendix~\ref{A-ZnM}:
\bea
Z_\zeta(|\vec p|) &=& \left[\intpfour A_s(|\vec{p}|,p_4)\right]^{-1}\nonumber\\[2mm]
\alpha(|\vec{p}|) &=& \frac{\intpfour A_t(|\vec{p}|,p_4)}{\intpfour A_s(|\vec{p}|,p_4)}
\nonumber\\[2mm]
M(|\vec p|) &=& \frac{\intpfour B_m(|\vec{p}|,p_4)}{\intpfour A_s(|\vec{p}|,p_4)}\,
\label{eq:stdress}
\eea
where the integrals here simply indicate a statistical average over the energy, 
given the $p_4$ independence of the dressing functions $A$'s and $B$
discussed in Sec.~\ref{sec:end_dip}.

To 
check renormalizability we now need to establish the scale invariance 
of $M(|\vec{p}|)$, $\alpha(|\vec{p}|)$ and study the scaling properties 
of $Z_\zeta(|\vec{p}|)$. 
Notice that $M$ and $Z_\zeta$ can also be extracted from 
the equal-times propagator (cf.~Eq.~(\ref{eq:decst})), while $\alpha$ 
always requires
the full $p_4$-dependent propagator (cf.~Eq.~(\ref{eq:fullprop})).

\begin{figure*}[htb]
\centering
\subfloat[$M$ from ($a$)-($g$).]{\label{scaling_M}\includegraphics[width=0.80\columnwidth,height=5truecm]{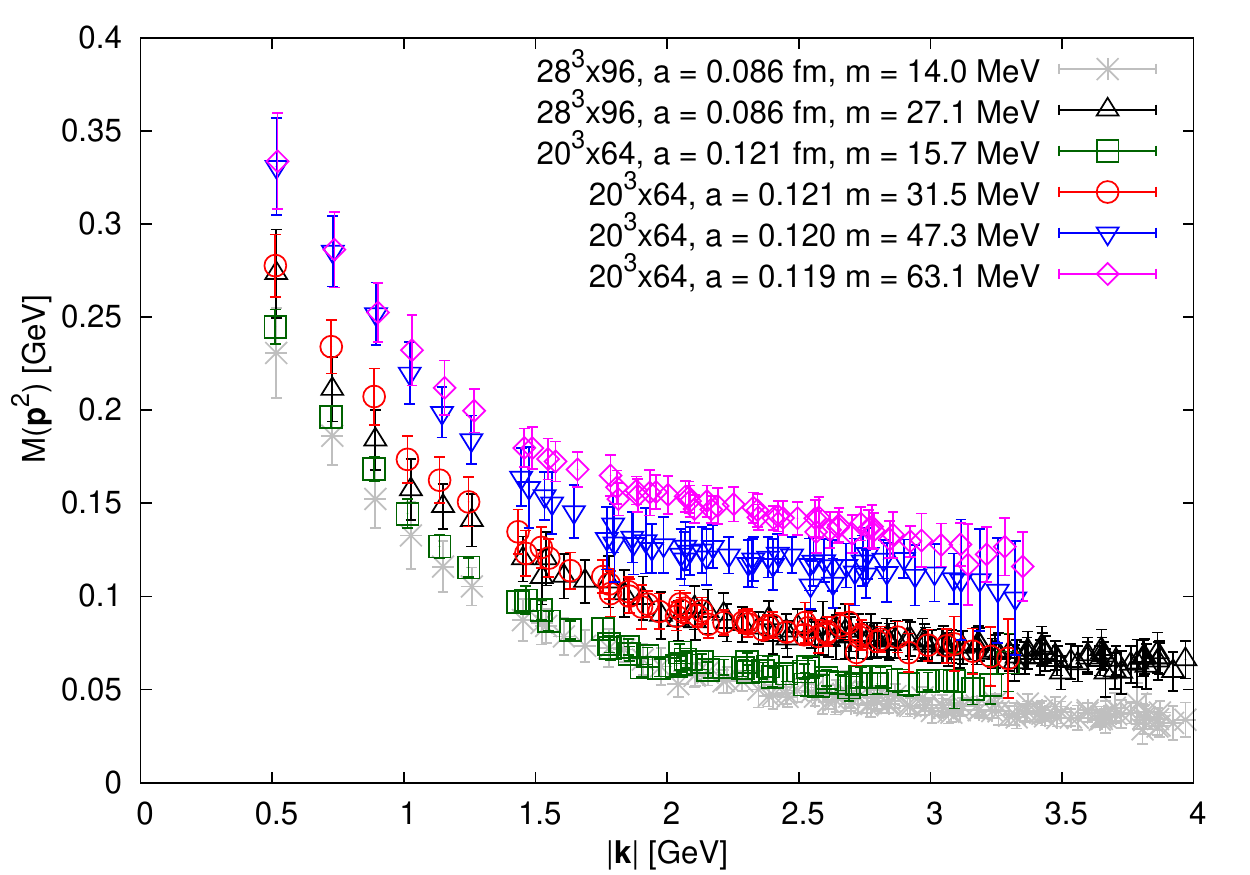}} \hspace*{1cm}
\subfloat[$M$ from ($f$) and ($i$)]{\label{scaling_large_M}\includegraphics[width=0.80\columnwidth,height=5truecm]{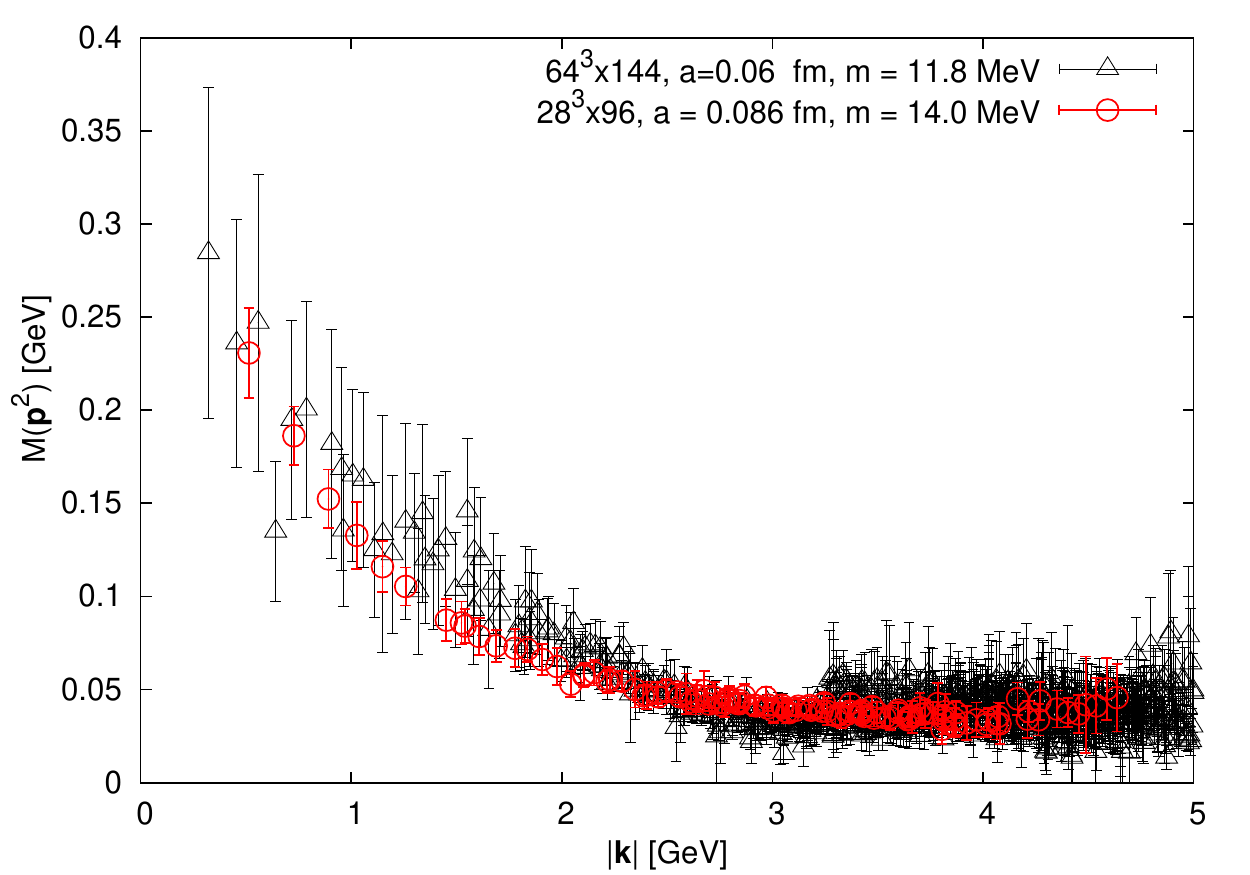}}\\
\subfloat[$M$ from ($f$) and ($h$)]{\label{unquenchingM}\includegraphics[width=0.80\columnwidth,height=5truecm]{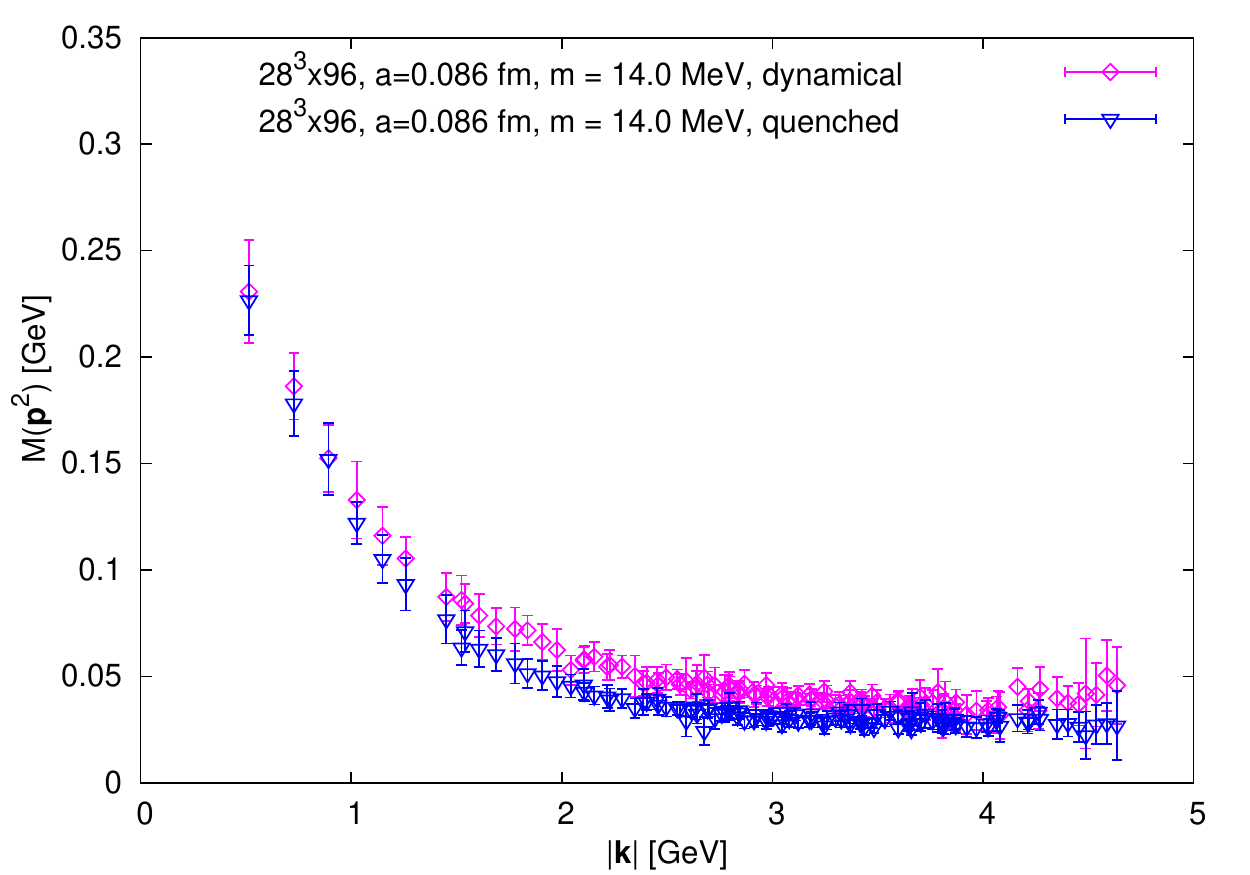}}  \hspace*{0.8cm}
\subfloat[$M$ from ($j$)-($l$).]{\label{quenchedM}\includegraphics[width=0.80\columnwidth,height=5truecm]{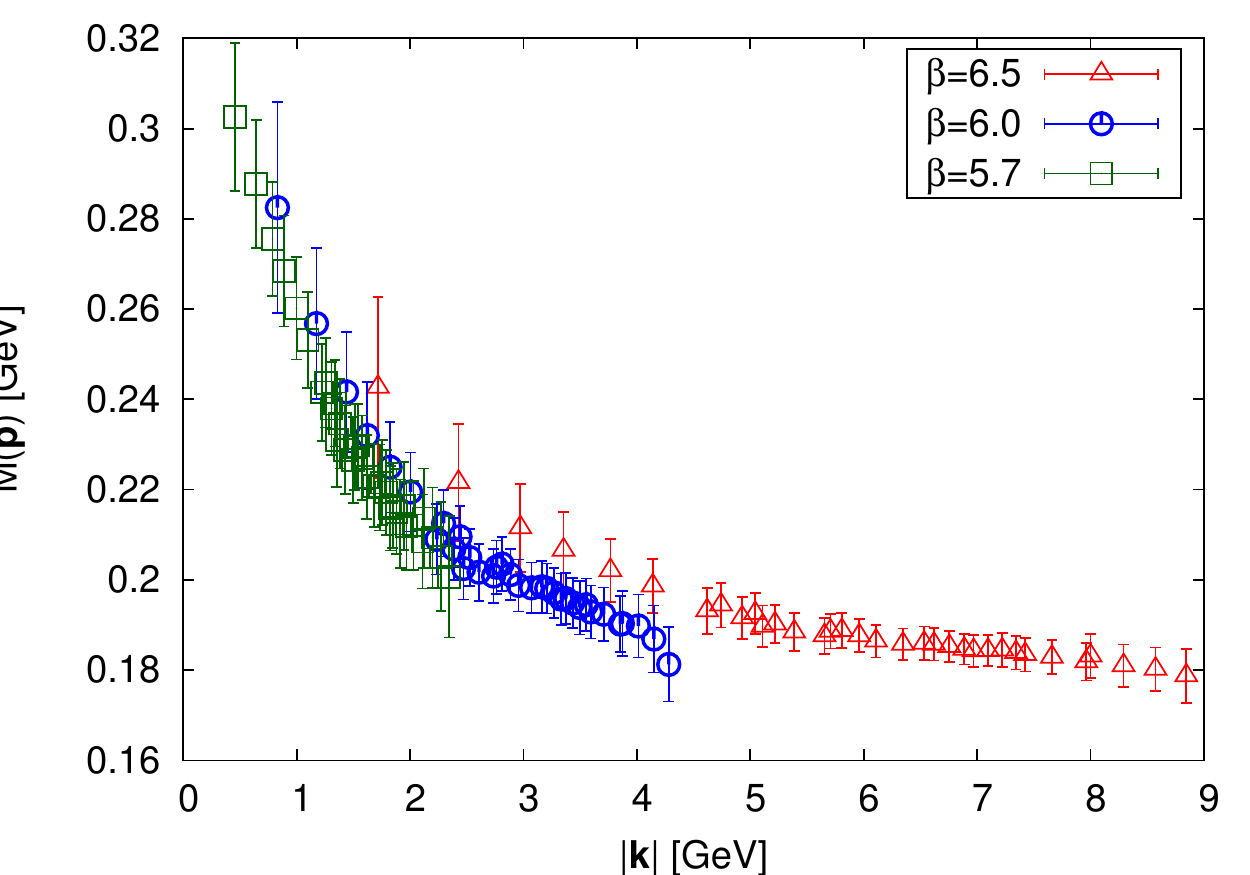}}
\caption{Scaling of the mass function $M$}
\label{2064_M}
\end{figure*}

\begin{figure*}[htb]
\centering
\subfloat[$Z$ from ($a$)-($g$).]{\label{unquenchingZall}\includegraphics[width=0.8\columnwidth,height=5truecm]{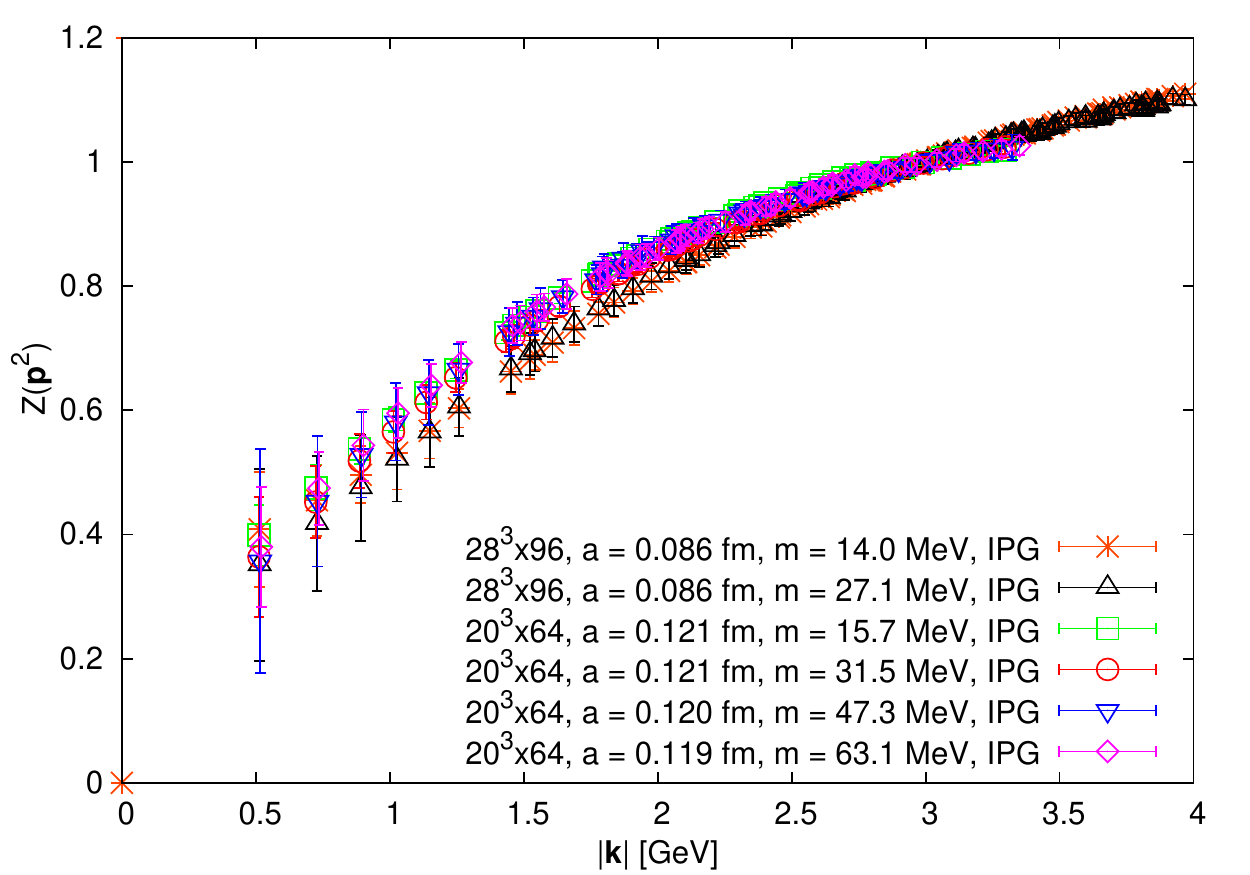}}
\hspace*{1cm}
\subfloat[$Z$ from ($b$) and ($g$)]{\label{scaling_Z}\includegraphics[width=0.8\columnwidth,height=5truecm]{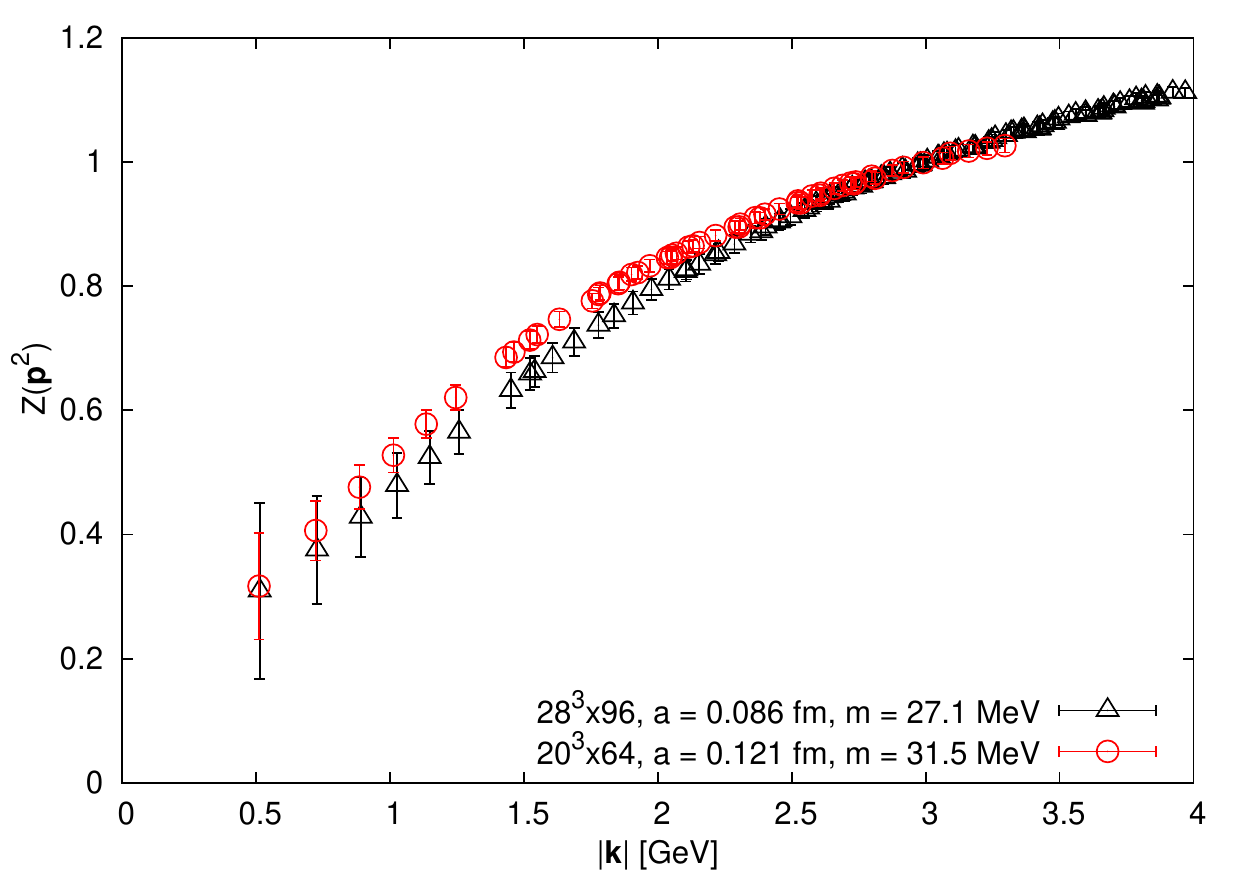}}\\
\subfloat[$Z$ from ($f$) and ($i$)]{\label{scaling_large_Z}\includegraphics[width=0.8\columnwidth,height=5truecm]{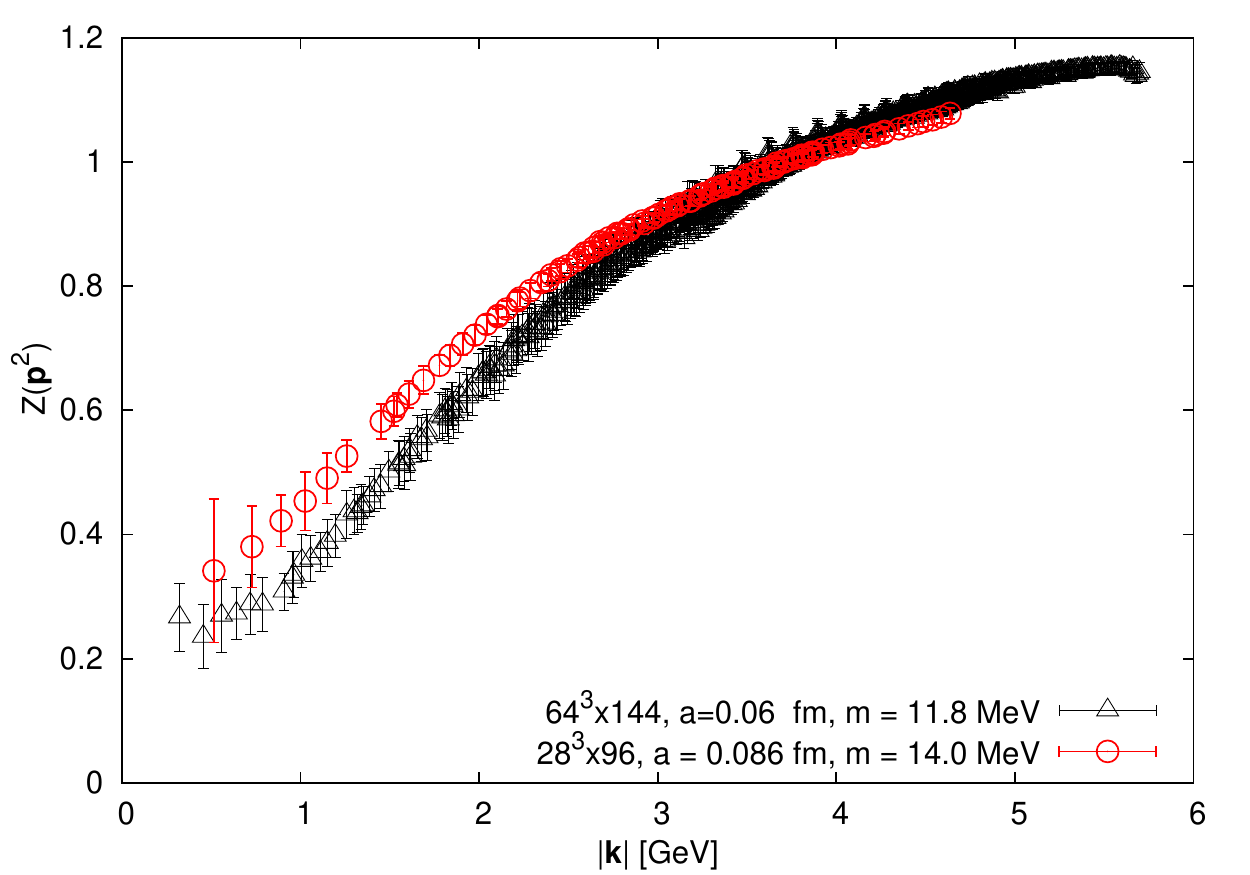}}
\hspace*{1cm}
\subfloat[$Z$ from ($f$)-($h$).]{\label{unquenchingZ}\includegraphics[width=0.8\columnwidth,height=5truecm]{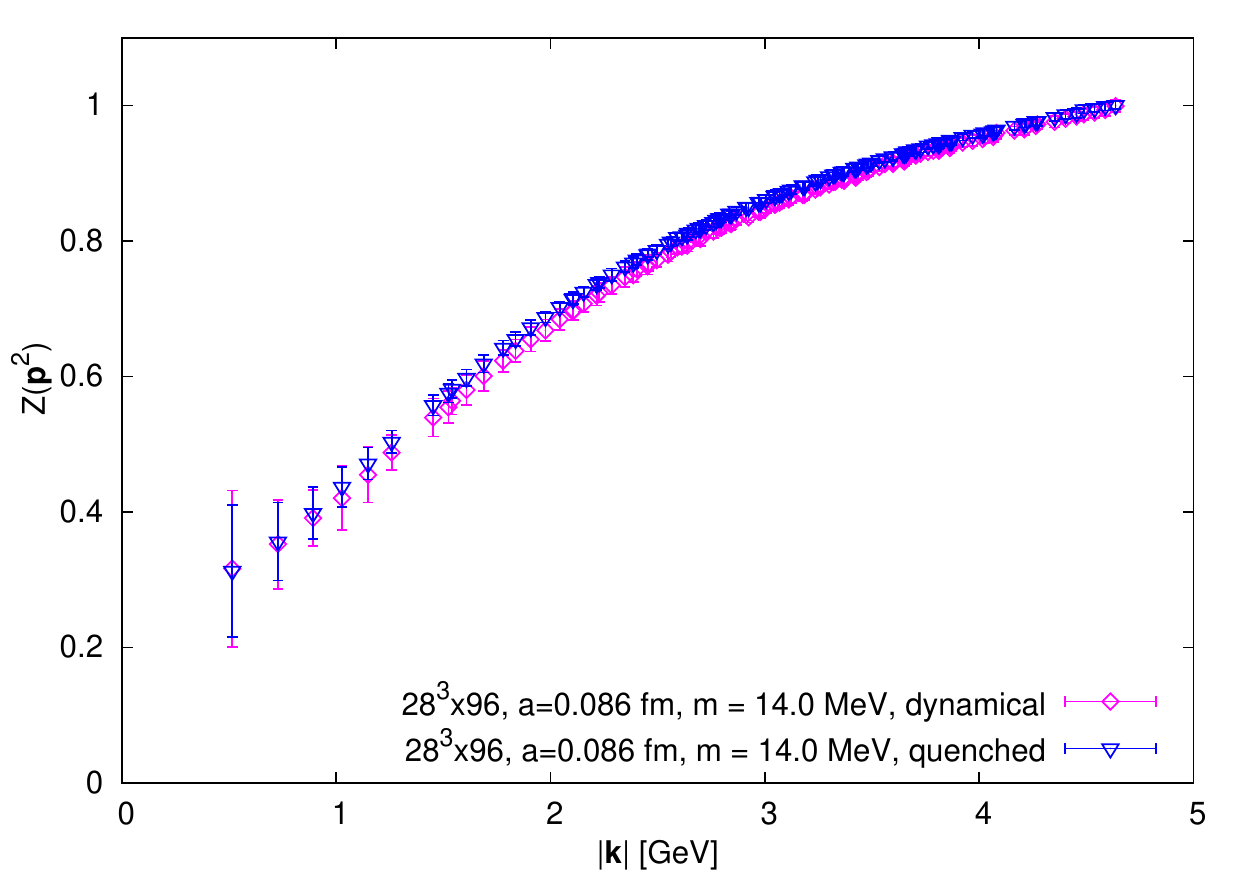}}
\caption{Scaling function $Z$. The renormalization point $\zeta$ is set at
\unit[3]{GeV} in (\ref{unquenchingZall}) and (\ref{scaling_Z}), 
\unit[3.7]{GeV} in \ref{scaling_large_Z} and \unit[4.64]{GeV} in 
\ref{unquenchingZ}}
\label{2064_Z}
\end{figure*}

\begin{figure*}[htb]
\centering
\subfloat[$A_t$ from ($a$)]{\label{2064_al_b}\includegraphics[width=0.8\columnwidth,height=5truecm]{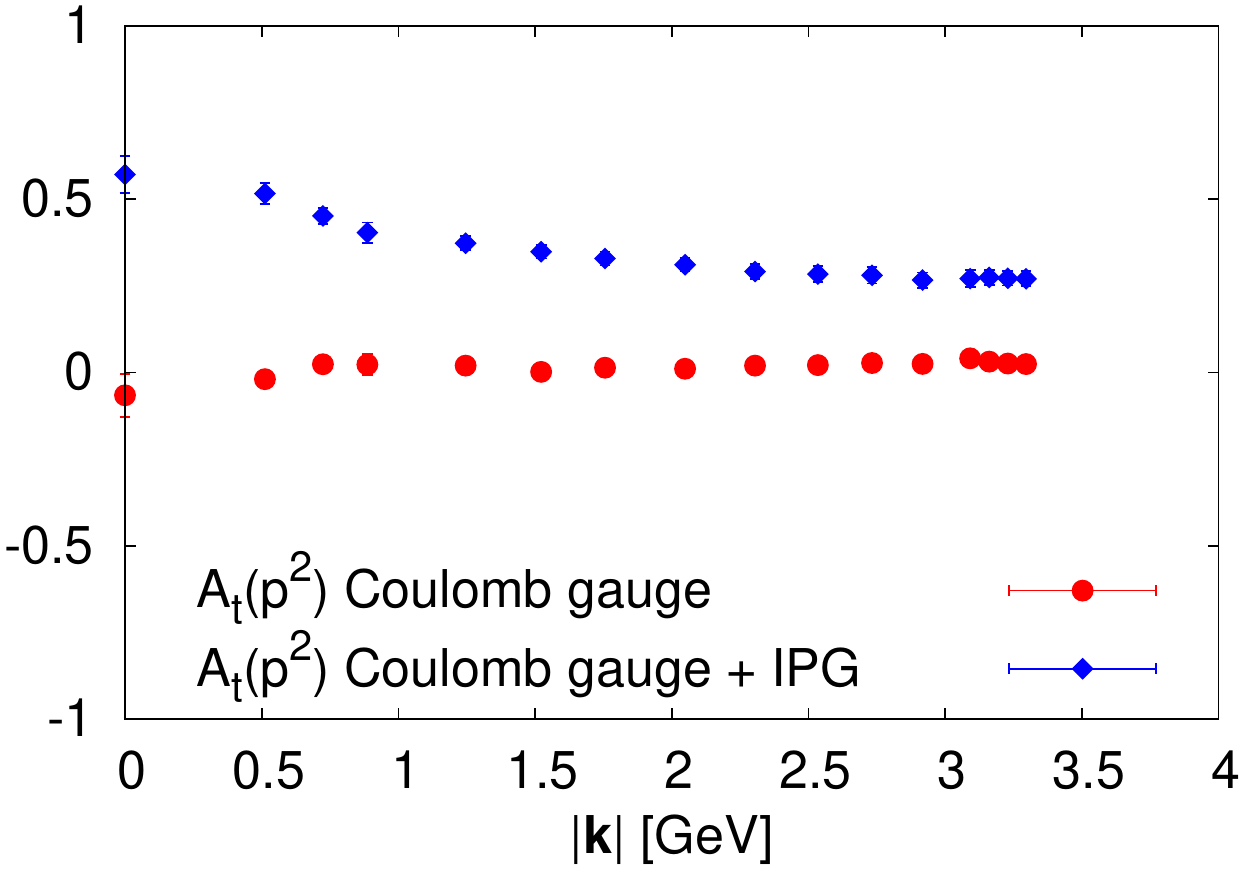}}
\hspace*{1cm}
\subfloat[$\alpha$ from ($a$)-($h$)]{\label{2064_al_d}\includegraphics[width=0.8\columnwidth,height=5truecm]{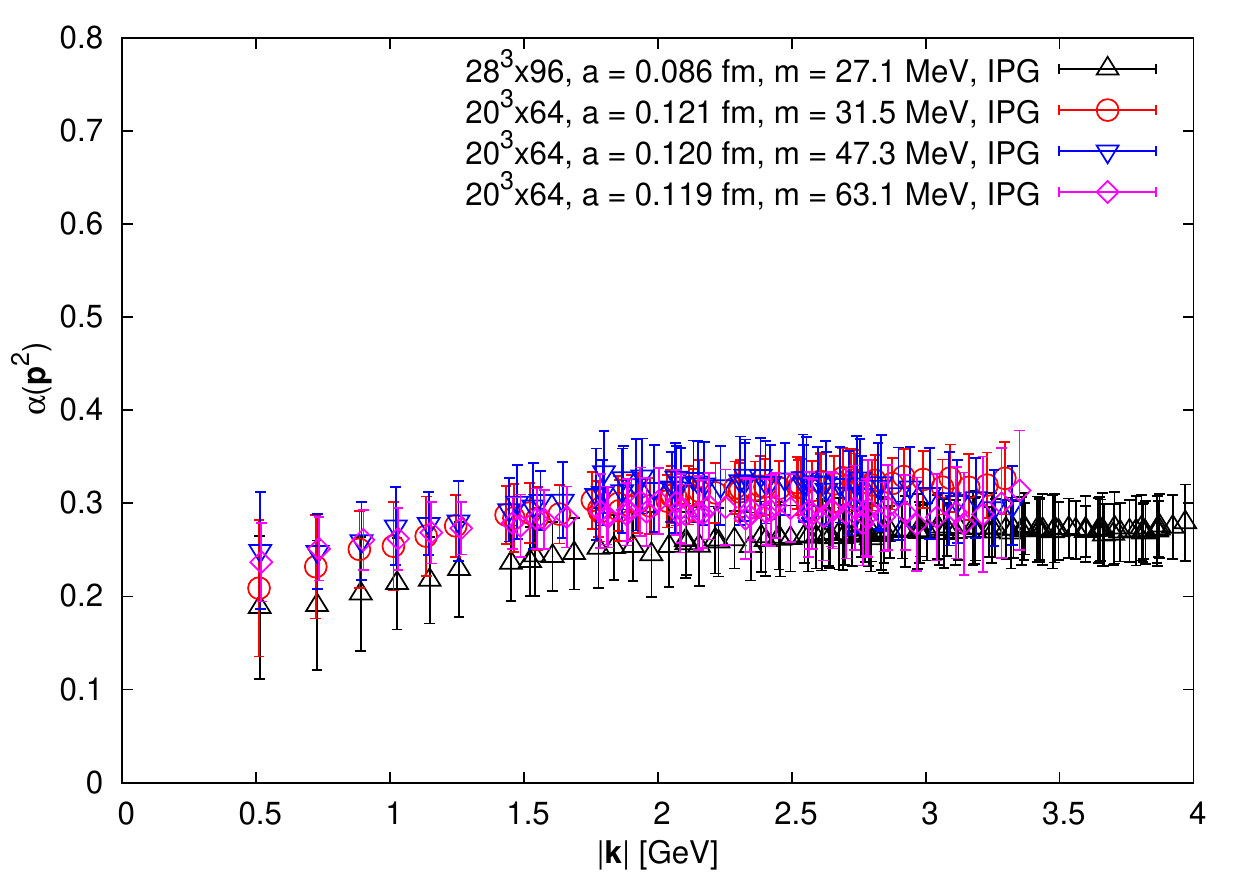}}\\
\hspace*{5mm}
\subfloat[$\alpha$ from ($e$), ($h$)]{\label{2064_al_c}\includegraphics[width=0.8\columnwidth,height=5truecm]{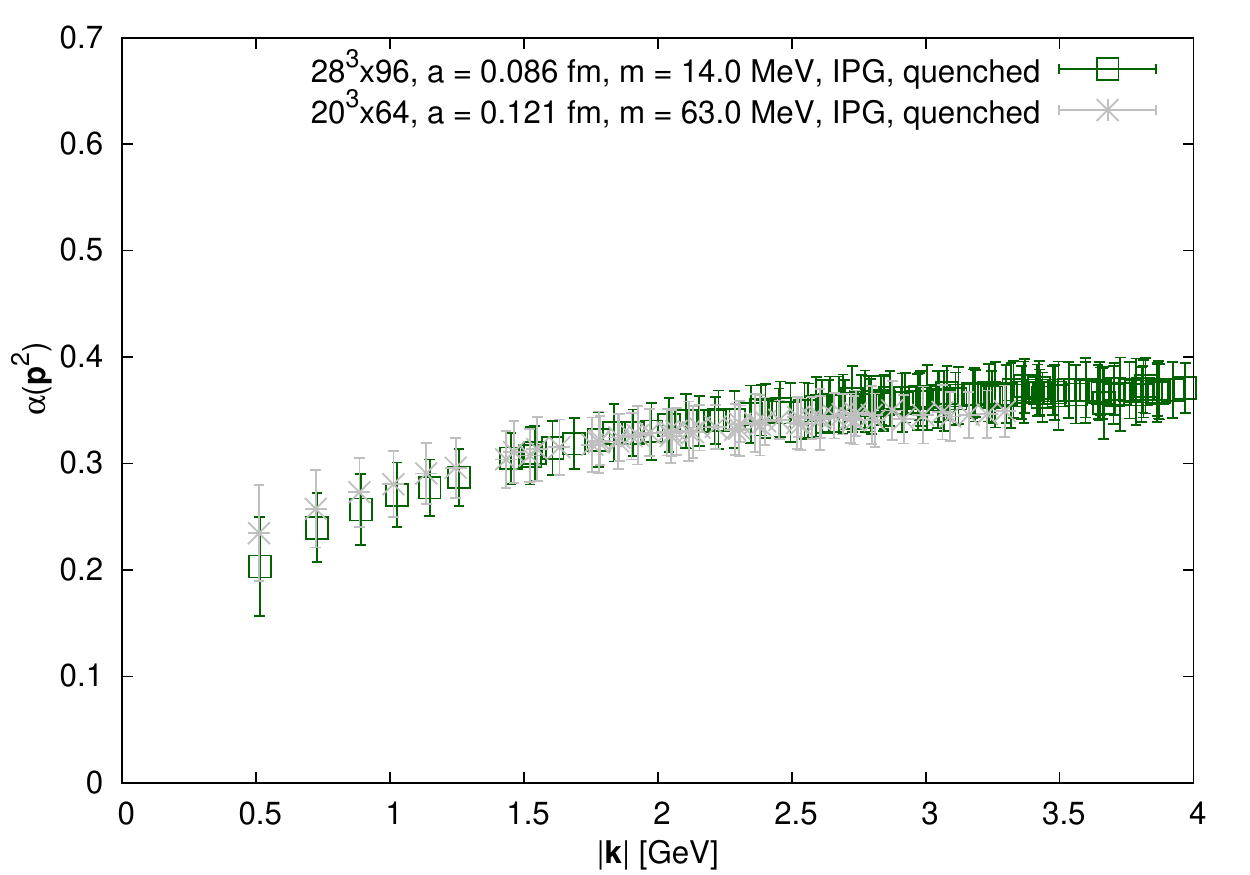}}
\hspace*{0.5cm}
\subfloat[$\omega_F$ from ($d$)-($e$)]{\label{2064_al_a}\includegraphics[width=0.9\columnwidth,height=5truecm]{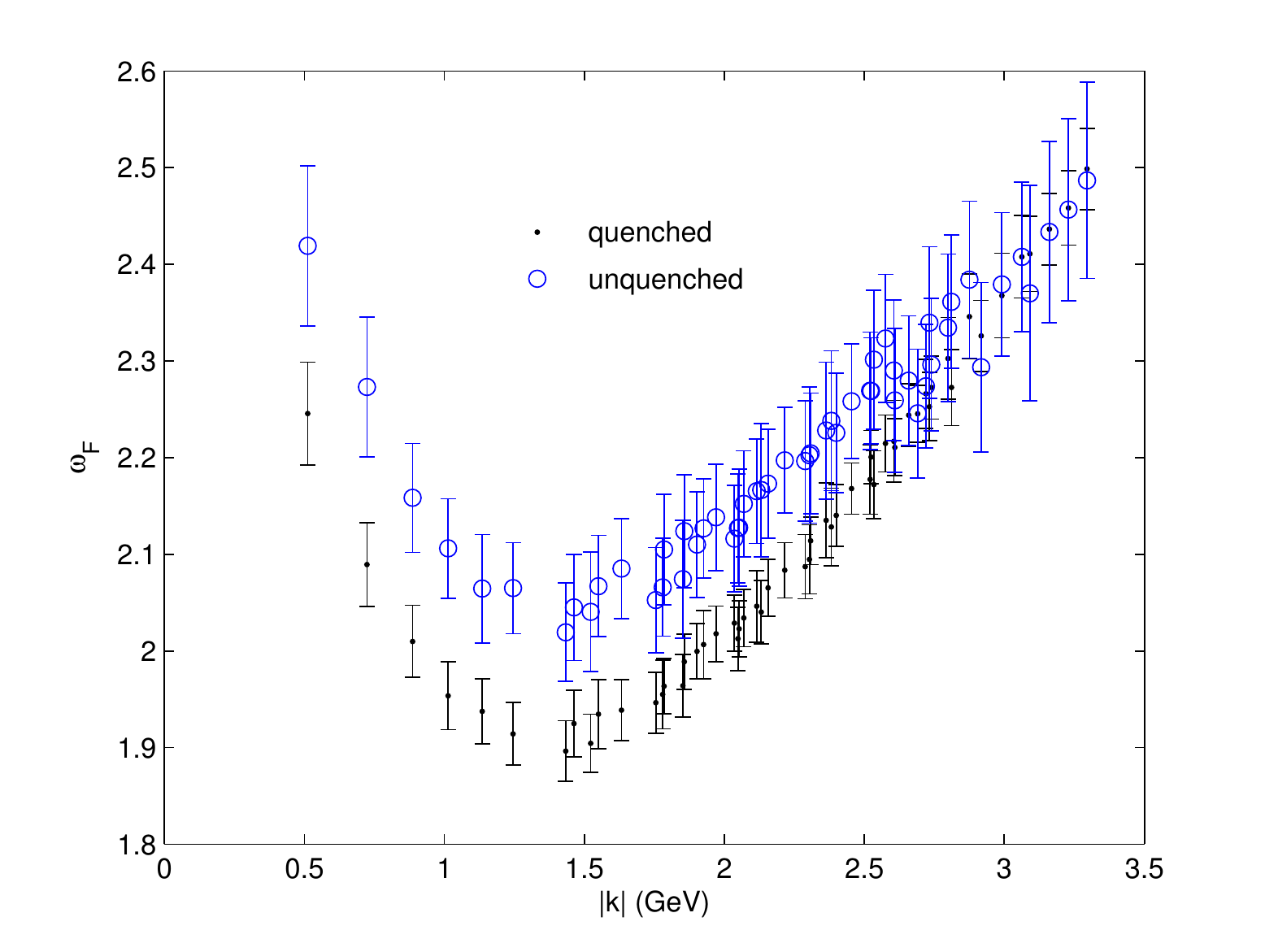}}
\caption{Scaling of the temporal function $\alpha$; $\omega_F$ as in Eq.~(\ref{eq:effene}).}
\label{2064_alpha}
\end{figure*}

In Fig.~\ref{scaling_M} we show the mass function $M(|{\vec p}|)$ from 
configuration sets ($a$)--($d$), with scale $a\approx\unit[0.12]{fm}$, 
compared to configuration sets ($f$) and ($g$),
which have a scale $a=\unit[0.086]{fm}$. 
These sets are chosen to have approximately the same physical volume,
so as to minimize finite size effects at the two different cutoffs.
As can be seen $M(|{\vec p}|)$ nicely agrees for the sets with similar 
masses, namely ($a$) ($\unit[15.7]{MeV}$) and ($f$) ($\unit[14.0]{MeV}$) on
one side and ($b$) ($\unit[31.5]{MeV}$) and ($g$) ($\unit[27.1]{MeV}$) on the 
other side.
We compare thus in Fig.~\ref{scaling_Z} the corresponding
wavefunction renormalization functions $Z_\zeta(|{\vec p}|)$ from configuration sets 
($g$) and ($b$),
finding a good agreement once we normalize both to $Z_\zeta(\zeta)=1$ for 
the scale $\zeta = \unit[3.0]{GeV}$.
The scaling of the Landau gauge quark propagator was checked on the same 
lattices in Ref. \cite{Parappilly:2005ei}, giving very similar results.

In Figs.~\ref{scaling_large_M}~and~\ref{scaling_large_Z} we compare 
$M(|{\vec p}|)$ and  $Z_\zeta(|{\vec p}|)$ from configuration sets ($f$) 
($\unit[14.0]{MeV}$) and ($i$) ($\unit[11.9]{MeV}$). Although on the latter
we have less statistics and could only calculate Kogut--Susskind fermions,
the agreement is still quite good.

We can thus conclude that the static propagator Eq.~(\ref{eq:decst}) is
multiplicatively renormalizable. It should be stressed, however, that
the wave function normalization $Z_\zeta$ (and $\alpha$, cf.~next paragraph)
show satisfactory scaling only when the improved Asqtad action is taken. This is
in contrast to the running mass $M(|\vec{p}|)$, which is very robust
against lattice artifacts, cf.~e.g.~Fig.~\ref{quenchedM}, where 
the unimproved Wilson action already gives a reasonable result ($m =212$ MeV 
for all three data sets). A similar effect
was also noticed in Landau gauge 
\cite{Bowman:2002bm,Parappilly:2005ei,Bowman:2001xh,Bowman:2002kn,Bowman:2005vx}.

Turning now to the full propagator Eq.~(\ref{eq:renns}),
Figs.~\ref{2064_al_d}-\ref{2064_al_c} show the function $\alpha$, i.e. 
the ratio of $A_t$ to $A_s$, for different configuration sets. The scaling 
behaviour is very good both for the dynamical configurations 
Fig.~\ref{2064_al_d} and for the quenched configurations  
Fig.~\ref{2064_al_c}. We can thus conclude that the full ($p_4$-dependent) 
propagator Eq.~(\ref{eq:renns}) is also multiplicatively renormalizable.
As argued above, this fact allows us to define a dispersion relation for a 
single pseudo-quark as in Eq.~(\ref{eq:effene}).

Interestingly, $Z_\zeta$ turns out to be much more suppressed in the IR than 
its Landau gauge counterpart, cf. Fig.~\ref{CvsLZ}, while $\alpha$ has a mild
momentum dependence. The resulting dispersion relation $\omega_F(|\vec{p}|)$ 
is then found to be \emph{IR enhanced}, as can be seen in Fig.~\ref{2064_al_a} 
for sets ($d$)-($e$), for which we have the best signal to noise ratio. 
If $Z_\zeta^2/\alpha$ should indeed prove to vanish in the IR according to 
a power law, 
\[
Z_\zeta^2(|\vec{p}|)/\alpha(|\vec{p}|) \propto |\vec{p}|^k
\]
we would find for the quark dispersion relation a behaviour similar to the 
Gribov formula for the gluon \cite{Gribov:1977wm,Burgio:2008jr,Burgio:2008yg}: in the 
ultraviolet $\omega_F \propto |\vec{p}|$ and in the infra-read a 
power law $\omega_F \propto |\vec{p}|^{-k}$. The last relation describes
an infrared diverging effective (pseudo-)energy for the quark, which would 
extend the Gribov-Zwanziger confinement scenario to the fermionic sector
of QCD. With our present 
data we find a behaviour compatible with an infrared exponent of 
$k \approx 0.25$, but of course a reliable estimate would need better
statistics on larger volumes, which we plan to collect in the
near future. Notice also that the minimum of $\omega_F(|\vec{p}|)$ is around
$|\vec{k}|\simeq 1.2$ GeV, which is compatible with the $a_t \to 0$ 
extrapolation of the gluonic Gribov mass
\cite{Burgio:2012bk} and the estimate of $\Lambda$ we will extract in 
Sec.~\ref{sec:chir}.

\subsection{Effects of dynamical quarks}
\label{sec:unquenching}

The dynamical and quenched configuration sets ($f$) and ($h$) have same size 
and scale. This makes them well suited to study quantitatively
the effects of quenching. To this end we have calculated on both sets the 
Coulomb gauge quark propagator with equal bare mass, viz.~the light sea quark 
mass of set ($f$). The resulting mass and renormalization functions 
$M(|{\vec p}|)$ and $Z_\zeta(|{\vec p}|)$ are compared in 
Figs.~\ref{unquenchingM} and \ref{unquenchingZ}. We find a slight screening 
of the dynamical mass 
generation in the case of full QCD, similarly to what was seen in 
Landau gauge \cite{Bowman:2005vx}.

\subsection{Comparison to Landau gauge}
\label{sec:lan}

\begin{figure*}[htb]
\centering
\subfloat[$M_{Coul.}$ vs. $M_{Lan.}$ from ($g$)]{\label{CvsLM}\includegraphics[width=0.8\columnwidth,height=5truecm]{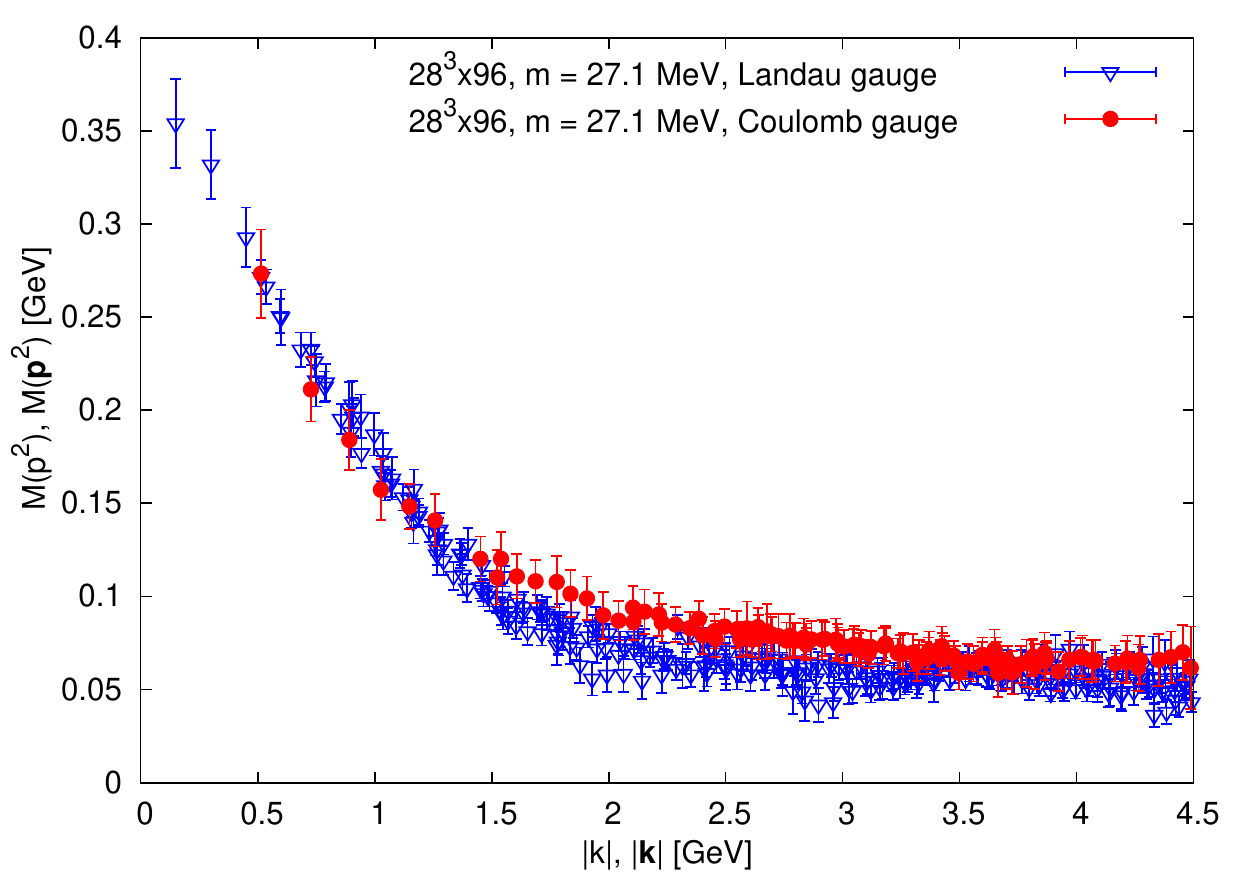}}
\hspace*{1cm}
\subfloat[$Z_{Coul.}$ vs. $Z_{Lan.}$ from ($g$)]{\label{CvsLZ}\includegraphics[width=0.8\columnwidth,height=5truecm]{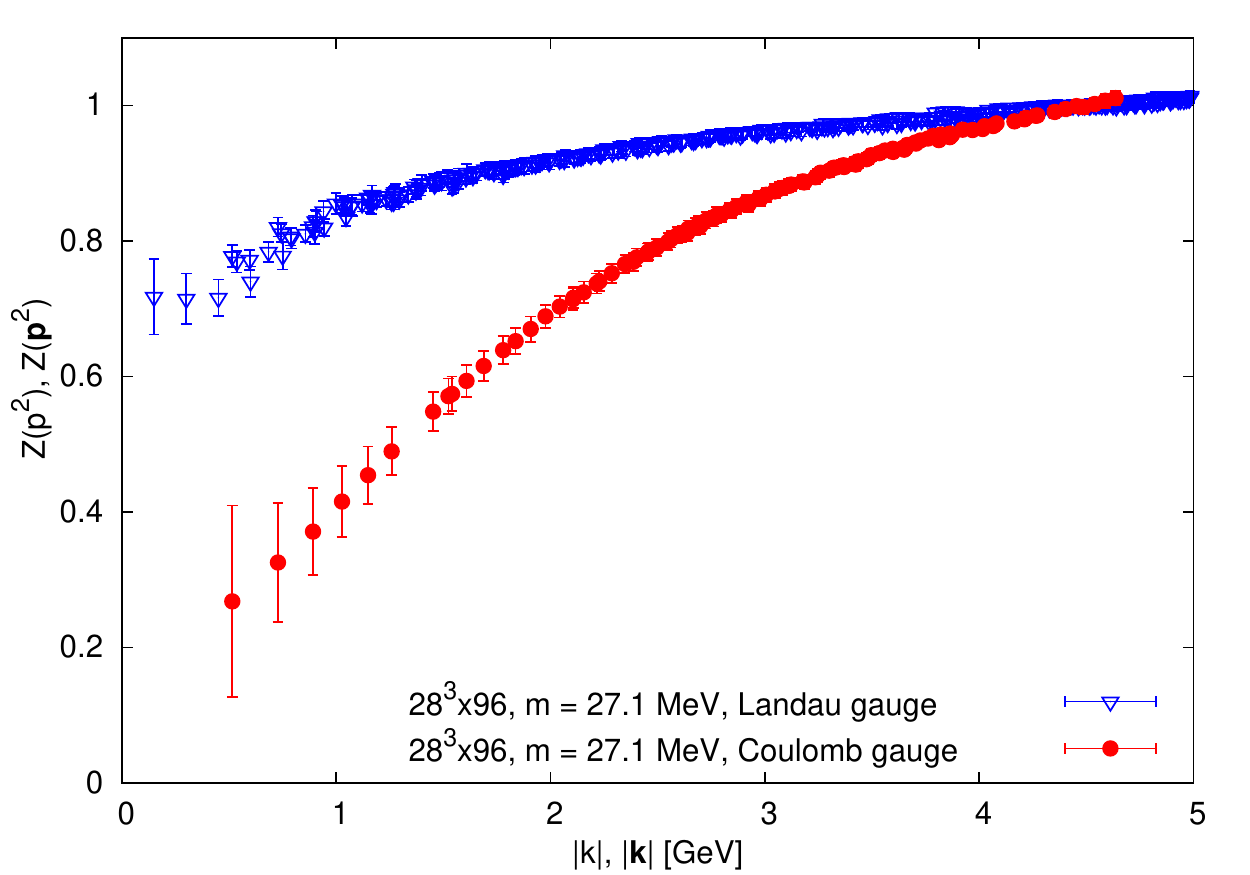}}\\
\subfloat[$M$ from ($a$)--($d$)]{\label{chirallimit_small}\includegraphics[width=0.8\columnwidth,height=5truecm]{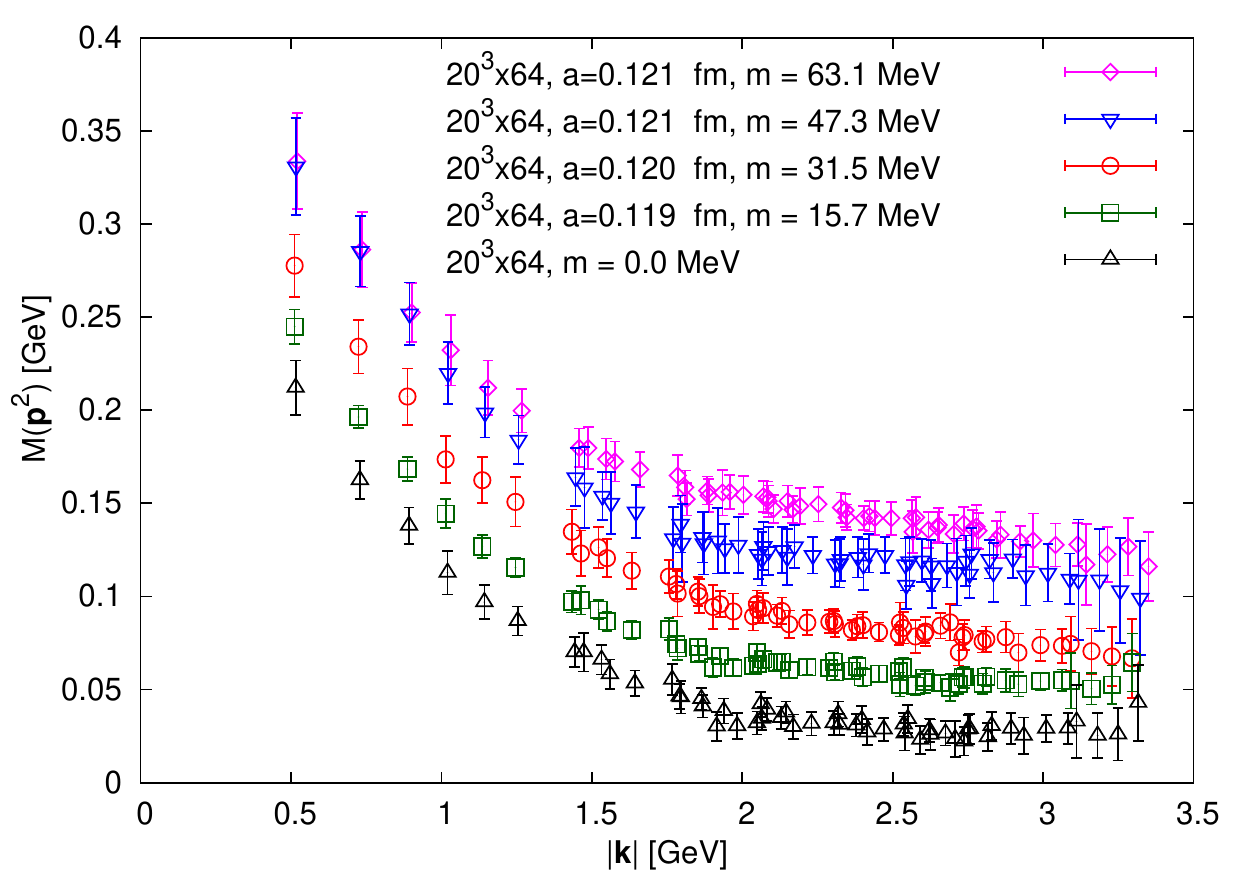}}
\hspace*{1cm}
\subfloat[$M$ from ($i$)]{\label{chirallimit_large}\includegraphics[width=0.8\columnwidth,height=5truecm]{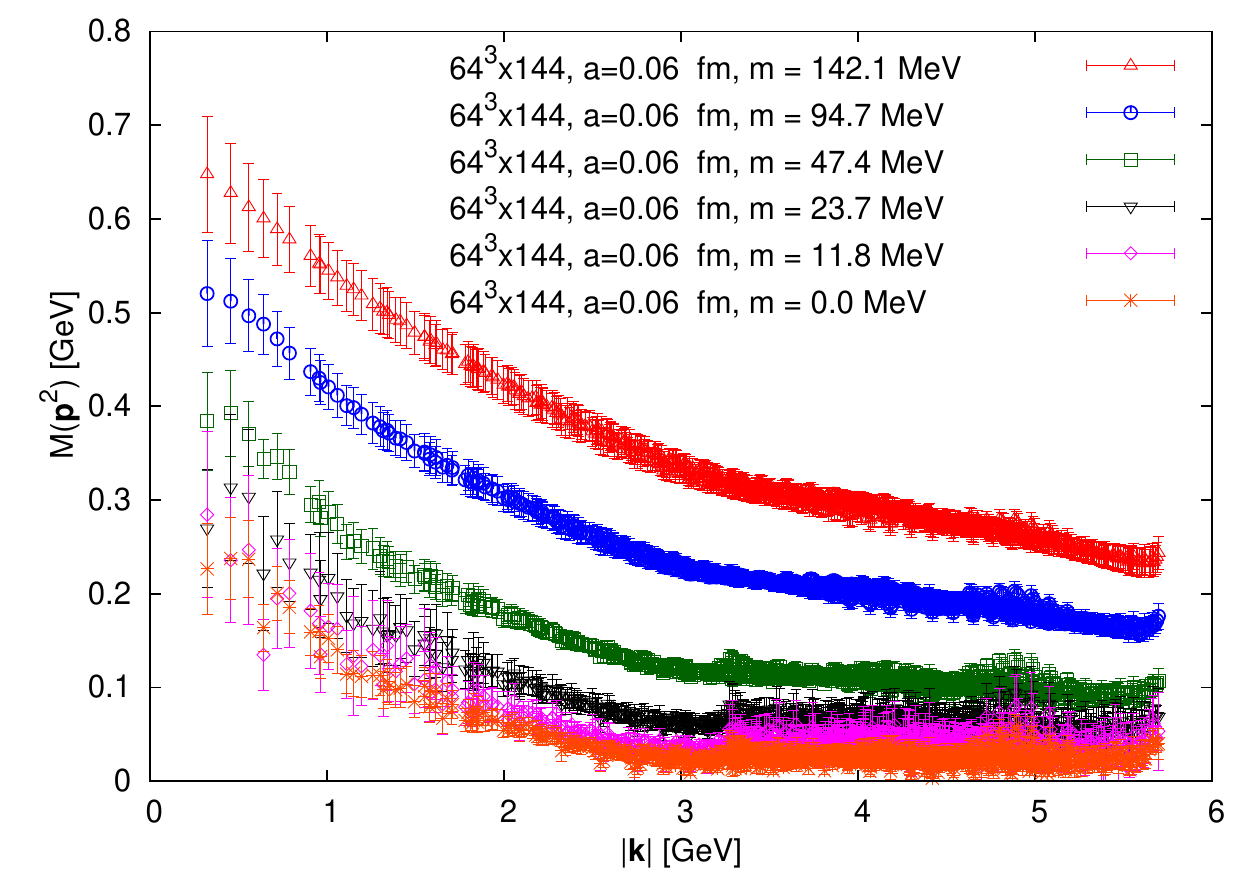}}
\caption{Landau vs. Coulomb gauge (top); chiral behaviour of $M$ (bottom)}
\label{2064_mix}
\end{figure*}

In Fig.~\ref{CvsLM} and \ref{CvsLZ} we show the comparison between the
mass and renormalization function in Coulomb gauge, Eq.~(\ref{eq:decst}),
and Landau gauge, Eq.~(\ref{eq:dec}). The data for the Landau case was taken 
from Ref.~\cite{Parappilly:2005ei}.
Up to slight differences in the intermediate and high momentum region,
which are however still within our error bars, $M$ is 
almost identical in both gauges, especially in the IR, while $Z_\zeta$ shows a 
much stronger 
gauge dependence. For this comparison, the renormalization point
has been set at $\zeta = $ \unit[4.64]{GeV}. 
These findings nicely agree with the discussion in Sec.~\ref{sec:ren}.

\subsection{Chiral limit}
\label{sec:chir}

\begin{figure*}[htb]
\centering
\subfloat[]{\label{ch1}\includegraphics[width=0.8\columnwidth,height=5truecm]{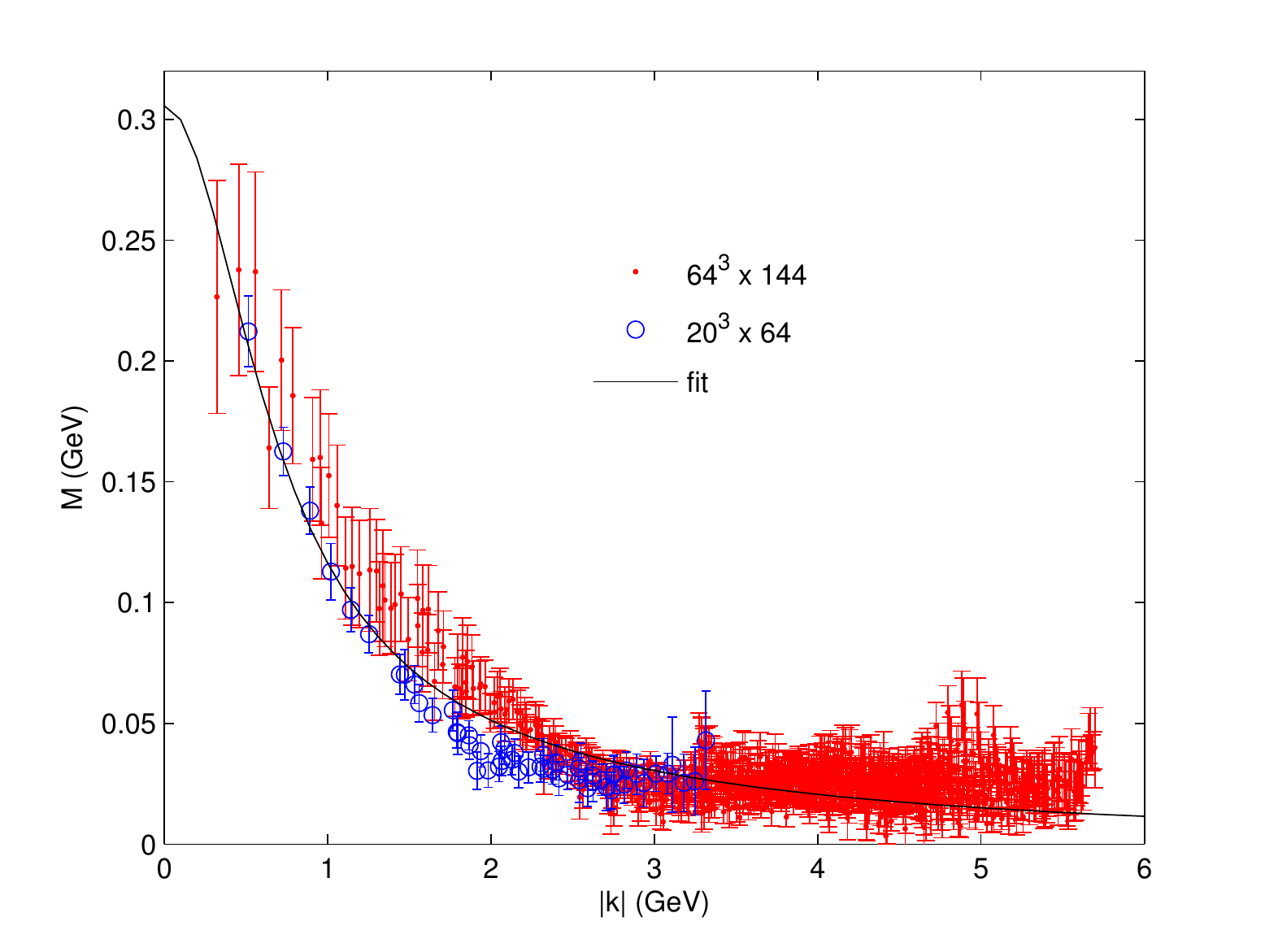}}
\hspace*{1cm}
\subfloat[]{\label{ch2}\includegraphics[width=0.8\columnwidth,height=5truecm]{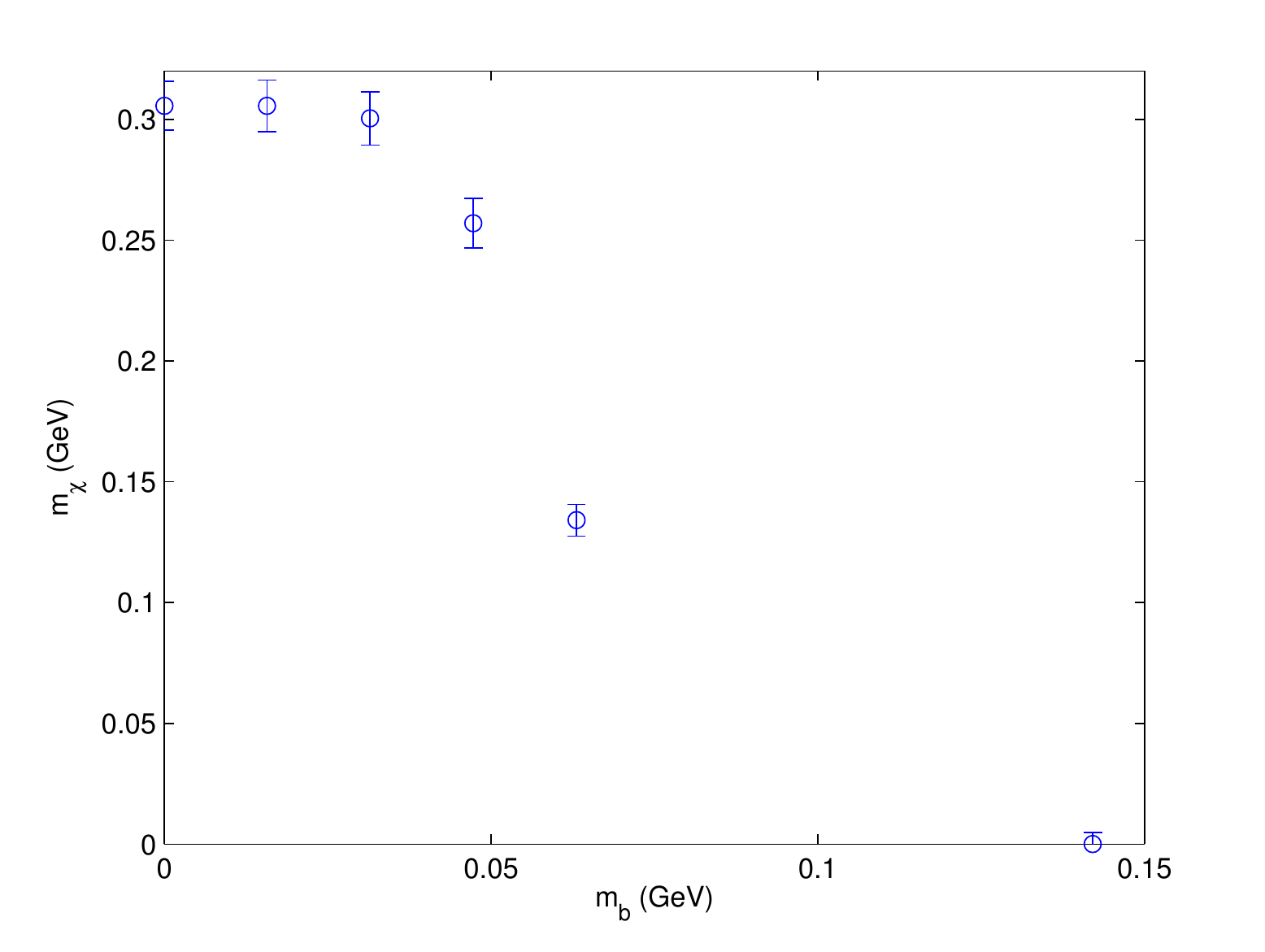}}\\
\subfloat[]{\label{ch3}\includegraphics[width=0.8\columnwidth,height=5truecm]{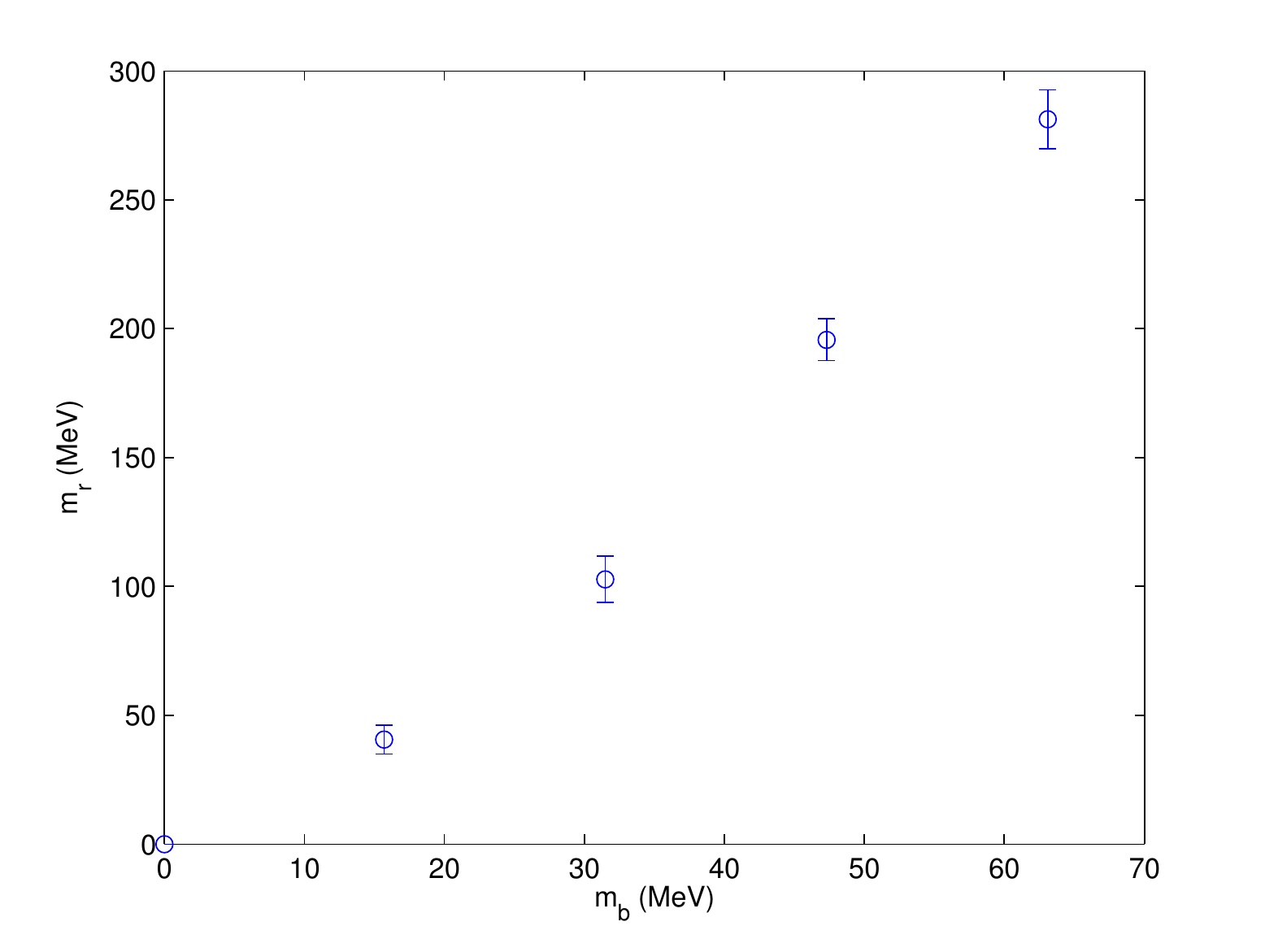}}
\hspace*{1cm}
\subfloat[]{\label{ch4}\includegraphics[width=0.8\columnwidth,height=5truecm]{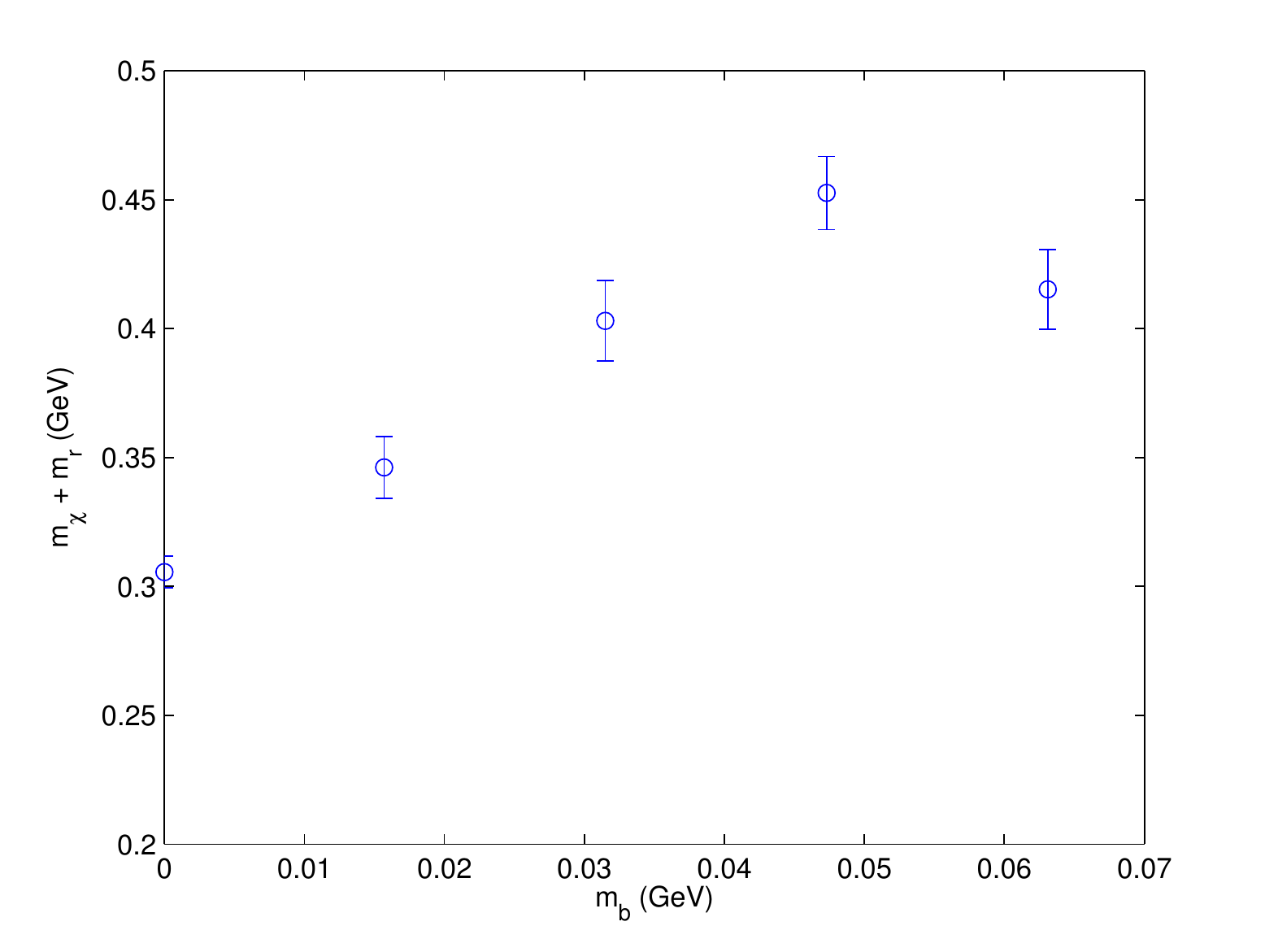}}
	\caption{The chiral limit taken from the dynamical point of 
        configuration sets ($a$)--($d$), compared to the chiral limit from 
set ($i$) (partially quenched) and to the fit to Eq.~(\ref{eq:ch}).}
\label{chirallimit}
\end{figure*}

For the configuration sets ($a$)--($d$) the dynamical masses decrease at 
constant cutoff. We have therefore attempted a chiral extrapolation of the 
mass function determined for these sets with Asqtad fermions at the dynamical 
point. This procedure should minimize systematic errors due to partial 
quenching.
The result is shown in Fig.~\ref{chirallimit_small}. In addition, we have also 
made a chiral extrapolation of the set ($i$) using  Kogut-Susskind fermions
(Fig.~\ref{chirallimit_large}), 
where the mass has been fixed to five different values, with only the last at 
the dynamical point. The result of the latter, although quite noisy in the IR, 
allows us to extend the momentum range and extract further information in the 
UV. 
Within error bars the two chiral extrapolations agree, cf.~Fig.~\ref{ch1}. 
We find that all data can be well described by a function of the form:
\bea
M(|\vec{k}|,m_b) &=& \frac{m_\chi(m_b)}{1+b\, \frac{|\vec{k}|^2}{\Lambda^2}\,
\log{\left(e+\frac{|\vec{k}|^2}{\Lambda^2}\right)}^{-\gamma}}\nonumber\\ 
&+& 
\frac{m_r(m_b)}{\log{\left(e+\frac{|\vec{k}|^2}{\Lambda^2}\right)}^{\gamma}}
\label{eq:ch}
\eea
where $b$ is a constant,  $\gamma$ is the anomalous dimension and $\Lambda$ 
is the QCD scale. In this fit we have also defined a renormalized Coulomb 
dressed mass $m_r$ and a chiral quark mass, both of which depend 
on the bare mass $m_b$. Performing a simultaneous fit 
for all data we find 
$b = 2.9(1)$, 
$\gamma = 0.84(2)$ and $\Lambda = 1.22(6)$ GeV with $\chi^2$/d.o.f.$=1.06$. 
For the extrapolated data of 
Fig.~\ref{ch1} we further find (constraining $m_r \equiv 0$) $m_\chi = 0.31(1)$ 
GeV, in very good agreement
with the constituent mass expected from quark models. In Fig.~\ref{ch2} we 
show the dependence of $m_\chi$ on the bare mass $m_b$. For bare masses larger 
than $m_b \simeq 100$ MeV the chiral quark mass 
practically disappears; on the other hand we find that the Coulomb gauge 
dressed mass $m_r$ increases more than linearly with $m_b$, cf.~Fig~\ref{ch3}.
One should not however over-interpret such result, since the splitting between 
the
two masses in the intermediate region will strongly depend on the fit function 
chosen.
A more reliable extraction of $m_r$ will need to be based on data at smaller
lattice spacing, pushing the fits in the higher UV region.
Finally, Fig.~\ref{ch4} displays the total constituent IR mass 
$M(0) = m_\chi + m_r$ from Eq.(\ref{eq:ch}) as a function of $m_b$; this should 
turn out to be fit independent as soon as the available data will push far 
enough in the IR region.

Similar calculations with Asqtad fermions at the finest spaced lattice and 
with the largest overall volume are currently under way to confirm our results.

\section{Conclusions}
\label{sec:conc}
In this paper we have studied the static and full quark propagator in Coulomb 
gauge. We have shown that, for improved actions minimizing discretization
errors, $S(p)$ has a trivial energy dependence and therefore
both $S(p)$ and the equal-time propagator $S(|\vec{p}|)$ are multiplicatively
renormalizable. The mass function entering both propagators agrees 
semi-quantitatively with its Landau gauge counterpart. However, for future
work, it will allow for a 
more efficient determination of the chiral parameters, the constituent quark 
masses and the quark anomalous dimension, since as we have shown the Dirac 
operator in Coulomb gauge is energy independent and its static version,
which holds all information about the running mass and the wave function 
renormalization, can be inverted at each fixed time slice.

Moreover, the full (non-static) Coulomb propagator gives access to an 
effective energy for the quark field. With our current data, the dispersion
relation is compatible with a Gribov kind of behaviour. Quark confinement 
could thus be explained in this picture by a diverging IR effective energy, 
which is a first evidence that the Gribov-Zwanziger confinement scenario could 
be extended to the fermionic sector of QCD.

%--------------
% Acknowledgments:
%--------------

\begin{acknowledgments}
We wish to thank P. O. Bowman for giving us access to the Landau gauge data 
used in Sec.~\ref{sec:lan} and for careful reading of the manuscript. 

This work was partially 
supported by DFG grant Re864/6-3; M.S. is supported by the Research Executive 
Agency (REA) of the European Union under Grant Agreement 
PITN-GA-2009-238353 (ITN STRONGnet).
\end{acknowledgments}

%--------------
% Appendix:
%--------------

\appendix

\section{Fixing the residual gauge}
\label{app:tgauge}

After fixing the Coulomb gauge at each time-slice $x_4$ we are still 
left with the freedom to perform a global gauge transformation $g(x_4)$ at
each $x_4$. To determine it we implement the lattice equivalent to the 
residual continuum condition (\ref{eq:tgauge}). Let us begin by averaging, 
for every fixed time-slice $x_4$, over the spatial coordinates 
$\vec{x}$ in $U_4(x)$,
\be
\label{intpolygauge}
	\hat u(x_4) \equiv \frac{1}{L^3}\sum_{\vec x} U_4(x)\,.
\ee 
Next we project the matrix $\hat u(x_4)$ back to \SU{3}
by Cabibbo--Marinari cooling \cite{Cabibbo:1982zn}, i.e.~we look for 
a $u(x_4) \in SU(3)$ satisfying:
\be
\max_{u(x_4)}\Re\tr{u(x_4)\hat u^\dagger(x_4)}.
\label{eq:projectSU3}
\ee
Although such projection is not unique, we believe it to be a ``natural'' 
choice; of course, different projections will lead to different temporal gauges.
However, since all dressing functions in $S$ turn out to be energy 
independent, the choice of a different prescription should eventually not 
influence the results.

We now seek a gauge transformation $g(x_4)$ such that
\be
	u(x_4)\stackrel{!}{\rightarrow}\mathrm{const.}
	\label{eq:appmq}
\ee
This can be achieved defining the integrated Polyakov loop:
\be
	P \equiv\tr{\prod_{x_4}u(x_4)}
\ee
and choosing $g(x_4)$ such that for all $x_4$
\be\label{P1T}
	g(x_4)u(x_4)g^\dagger(x_4+1) = P^{\nicefrac{1}{T}},
\ee
where $T$ is the temporal extension of the lattice and $P^{\nicefrac{1}{T}}$ 
is the $T$-th root of $P$. To do so we (arbitrarily) choose $g(0)=\1$ and 
then determine all $g(x_4)$ recursively,
\be
	g^\dagger(x_4+1) = u^\dagger(x_4)g^\dagger(x_4)P^{\nicefrac{1}{T}}.
\ee
If we now gauge all links via the transformation $g(x_4)$ we have 
\begin{eqnarray}
	U'_4(\vec{x},x_4) &=& g(x_4)U_4(\vec{x},x_4)g^\dagger(x_4+1) \\[2mm]
U'_i(\vec{x},x_4)&=&g(x_4)U_i(\vec{x},x_4)g^\dagger(x_4), 
\label{eq:glg}
\end{eqnarray}
respectively. The new temporal links $U_4'$ obey Eq.~(\ref{eq:appmq}); if 
$u \sim \hat{u}$ as in the case of $SU(2)$, this condition would via
Eq.~(\ref{intpolygauge}) translate into $\partial_4 \sum_{\vec{x}} U_4'(x)$, 
which is the equivalent to  the continuum condition (\ref{eq:tgauge}).
For $G=SU(3)$, however, the sum of colour matrices is not proportional to 
an $SU(3)$ matrix, and we can only choose $\hat{u}(x_4)$ as close as possible 
to $u(x_4)$. The cooling step Eq.~(\ref{eq:projectSU3}) implements thus
the continuum condition Eq.~(\ref{eq:tgauge}) as much as possible for the color 
group $G=SU(3)$. From Eq.~(\ref{eq:glg}), we also see that the new 
transformation 
$g(x_4)$ acts as a global gauge transformation at each fixed time slice $x_4$, 
i.e.~the Coulomb gauge condition remains unaffected.

\section{Staggered quark propagators}
\label{staggDetails}

We extend here to Coulomb gauge the decomposition of the staggered quark 
propagator in Landau gauge performed in \cite{Bowman:2001sz, Bowman:2005vx}.
The Kogut--Susskind propagator reads at tree-level:
\be
S^{(0)}_{\alpha\beta}(q)^{-1} =i\sum_\mu(\bar\gamma_\mu)_{\alpha\beta}\sin(\qhat_\mu) +
\mhat_0\bar\delta_{\alpha\beta}
\ee
where $\alpha$, $\beta$ are staggered multi-indices, 
$\alpha=(\alpha_1,\hdots,\alpha_4)$, $\alpha_\mu\in\{0, 1\}$, while 
$\bar\delta_{\alpha\beta}\equiv\prod_\mu\delta_{\alpha_\mu\beta_\mu|\textrm{mod} 2}$ and
\be
(\bar\gamma_\mu)_{\alpha\beta} \equiv (-1)^{\alpha_\mu} \bar\delta_{\alpha+\theta(\mu),\beta}\;,
\ee
where the $\nu^{\mathrm{th}}$ component of the 4-vectors $\theta(\mu)$ is given
by: 
\be
\theta_\nu(\mu)=\begin{cases}1 &\textrm{if } \nu<\mu,\\ 0&\textrm{else}\,,
\end{cases}
\ee
giving the staggered Dirac algebra
\be
\left\{\bar\gamma_\mu,\bar\gamma_\nu\right\}_{\alpha\beta}=2\delta_{\mu\nu}
\bar\delta_{\alpha\beta}.
\ee
The discrete momenta $\qhat_\mu\equiv aq_\mu$ are restricted for staggered 
fermions to the inner half of the Brillouin zone, 
$\qhat_\mu\in \left(-\frac{\pi}{2},\frac{\pi}{2}\right]$,
and are related to the ordinary lattice momenta 
$p_\mu\in \left(-\pi,\pi\right]$ via
\be
\phat_\mu = \qhat_\mu + \rho_\mu \pi, \quad \rho_\mu =\{0, 1\}.
\ee
In the following we will use the common abbreviation 
$\hat k_\mu\equiv\sin\hat q_\mu$.

Summing over one of the propagator's multi-indices, 
$\sum_\beta=\sum_{\beta_1,\hdots, \beta_4=0}^1$, we define
\be\label{staggeredpropG}
G^{(0)}_\alpha(q)\equiv\sum_{\beta}S^{(0)}_{\alpha\beta}(q) 
=\sum_\beta \frac{-i(\bar\gamma_\mu)_{\alpha\beta}\hat k_\mu + \mhat_0
\bar\delta_{\alpha\beta}}{\hat k^2+\mhat_0^2}. 
\ee
In order to evaluate the r.h.s. of \Eq{staggeredpropG}, we note that
\be
\sum_\beta (\bar\gamma_\mu)_{\alpha\beta} \equiv \sum_\beta (-1)^{\alpha_\mu} 
\bar\delta_{\alpha+\theta(\mu),\beta}=(-1)^{\alpha_\mu}\,.
\ee
We now want to extract the dressing functions of the inverse Coulomb propagator:
\begin{multline}
G^{-1}_\alpha(q)\equiv i (-1)^{\alpha_i}\hat{k}_ia {A}_s(|\vec{q}|,q_4^2) 
+i (-1)^{\alpha_4}\hat{k}_4 a {A}_t(|\vec{q}|,q_4^2)\\
+i (-1)^{\alpha_i+\alpha_4}\hat{k}_i a {A}_d(|\vec{q}|,q_4^2) 
+ {B}_m(|\vec{q}|,q_4^2)\,.
\end{multline}
On the other hand, we can also express the propagator in terms of 
new dressing functions
$\mathcal{A}_s$, $\mathcal{A}_t$, $\mathcal{A}_d$ and $\mathcal{B}_m$:
\begin{align}
\label{eq:stprop}
&G_\alpha(q)\equiv -i (-1)^{\alpha_i}\hat{k}_ia \mathcal{A}_s(|\vec{q}|,q_4^2) 
-i (-1)^{\alpha_4}\hat{k}_4 a \mathcal{A}_t(|\vec{q}|,q_4^2)\nonumber\\
&-i (-1)^{\alpha_i+\alpha_4}\hat{k}_i a \mathcal{A}_d(|\vec{q}|,q_4^2) 
+ \mathcal{B}_m(|\vec{q}|,q_4^2)\nonumber\\
&= \big[-i(-1)^{\alpha_i}\hat{k}_iaA_s(|\vec{q}|,q_4^2) 
-i(-1)^{\alpha_4}\hat{k}_4aA_t(|\vec{q}|,q_4^2)\nonumber\\
&-i(-1)^{\alpha_i+\alpha_4}\hat{k}_i a A_d(|\vec{q}|,q_4^2) 
+ B_m(|\vec{q}|,q_4^2)\big]/D^2(|\vec{q}|,q_4^2)
\end{align}
where we have defined
\begin{multline}
D^2(|\vec{q}|,q_4^2)\equiv \sum_i\hat{k}_i^2a^2 A_s^2(|\vec{q}|,q_4^2)  
+ \hat{k}_4^2a^2A_t^2(|\vec{q}|,q_4^2) \\
+ \sum_i\hat{k}_i^2a^2A_d^2(|\vec{q}|,q_4^2) + B_m^2(|\vec{q}|,q_4^2)\,.
\end{multline}
After some algebra it can be shown that 
\be
\label{eq:inv}
D^2(|\vec{q}|,q_4^2) \,\mathcal{D}^2(|\vec{q}|,q_4^2) =1\,
\ee
where
\begin{multline}
\mathcal{D}^2(|\vec{q}|,q_4^2)\equiv \sum_i\hat{k}_i^2a^2 
\mathcal{A}_s^2(|\vec{q}|,q_4^2) +\hat{k}_4^2a^2\mathcal{A}_t^2(|\vec{q}|,q_4^2)\\
+ \sum_i\hat{k}_i^2a^2\mathcal{A}_d^2(|\vec{q}|,q_4^2) + 
\mathcal{B}_m^2(|\vec{q}|,q_4^2)\,.
\end{multline}
Now we multiply Eq.~(\ref{eq:stprop}) by 
$(-1)^{\alpha_i}$, $(-1)^{\alpha_4}$, $(-1)^{\alpha_i+\alpha_4}$ or 1, respectively, 
and sum over $\alpha$ using 
\be
\sum_{\alpha} (-1)^{\alpha_\mu+\alpha_\nu} = 16\,\delta_{\mu\nu}\,.
\ee
Taking finally the trace with respect to color indices we obtain:
\begin{multline}
\label{eq:as}
	\mathcal A_s(|\vec{q}|,q_4^2)\equiv\frac{A_s(|\vec{q}|,q_4^2)}
{D^2(|\vec{q}|,q_4^2)} 
		= \frac{i}{16N_c\sum_i\hat{k}_i^2a } \\
	\times\sum_i\sum_\alpha(-1)^{\alpha_i}\hat{k}_i\tr{G_\alpha(q)}, 
\end{multline}
\begin{multline}
\label{eq:at}
\mathcal A_t(|\vec{q}|,q_4^2)\equiv\frac{A_t(|\vec{q}|,q_4^2)}
{D^2(|\vec{q}|,q_4^2)} 
= \frac{i}{16N_c\hat{k}_4^2a }\\
\times \sum_\alpha(-1)^{\alpha_4}\hat{k}_4\tr{G_\alpha(q)}, 
\end{multline}
\begin{multline}
\label{eq:ad}
	\mathcal A_d(|\vec{q}|,q_4^2)\equiv\frac{A_d(|\vec{q}|,q_4^2)}
{D^2(|\vec{q}|,q_4^2)} = \frac{i}{16N_c\sum_i\hat{k}_i^2 a}\\ 
\times\sum_i\sum_\alpha(-1)^{\alpha_i+\alpha_4}\hat{k}_i\tr{G_\alpha(q)}, 
\end{multline}
\be
\label{eq:bm}
	\mathcal B_m(|\vec{q}|,q_4^2)\equiv\frac{B_m(|\vec{q}|,q_4^2)}
{D^2(|\vec{q}|,q_4^2)} 
		= \frac{1}{16N_c} \sum_\alpha\tr{G_\alpha(q)}.
\ee
From Eq.~(\ref{eq:inv}) and Eqs.~(\ref{eq:as}-\ref{eq:bm}) we can now easily
extract the dressing functions $A_s$, $A_t$, $A_d$ and $ B_m$.

\medskip
Turning next to Asqtad improved fermions, their tree-level propagator reads:
\begin{multline}\label{AsqtadDiracOp}
S^{(0)}_{\alpha\beta}(q)^{-1}=iu_0\sum_\mu(\bar\gamma_\mu)_{\alpha\beta}\sin(\qhat_\mu)\\
\times\left[ 1+\frac{1}{6} \sin^2(\qhat_\mu)\right] +\mhat\bar\delta_{\alpha\beta}
\end{multline}
and the decomposition given above can be performed in essentially the same way.
We only have to keep in mind, though, that the dressing functions will also get 
contributions from the tadpole factors $u_0$, which have to be eliminated 
a posteriori.

\section{Formulas for $Z$, $M$ and $\alpha$}
\label{A-ZnM}

To extract $Z(|\vec p|)$, $M(|\vec p|)$ and $\alpha$ we can proceed  as 
in Appendix~\ref{staggDetails}. Recalling that 
$A_d \equiv 0$ we write the propagator as ($k_\mu\equiv\sin p_\mu$):
\begin{align}
&S^{-1}(p) =i\veckslash a A_s(|\vec{p}|,p_4)+ 
i\kslash_4 a A_t(|\vec{p}|,p_4)+ B_m(|\vec{p}|,p_4)\\
&=  i\veckslash a \frac{\A_s(|\vec{p}|,p_4)}{\Dcal^2(|\vec{p}|,p_4)}
+i\kslash_4 a \frac{\A_t(|\vec{p}|,p_4)}{\Dcal^2(|\vec{p}|,p_4)} 
+ \frac{\B_m(|\vec{p}|,p_4)}{\Dcal^2(|\vec{p}|,p_4)}
\end{align}
and
\begin{align}
&S(p) = - i\veckslash a \A_s(|\vec{p}|,p_4) -i\kslash_4 a \A_t(|\vec{p}|,p_4)
+ \B_m(|\vec{p}|,p_4)\\
&= - i\veckslash a \frac{A_s(|\vec{p}|,p_4)}{D^2(|\vec{p}|,p_4)} 
-i\kslash_4 a \frac{A_t(|\vec{p}|,p_4)}{D^2(|\vec{p}|,p_4)} 
+ \frac{B_m(|\vec{p}|,p_4)}{D^2(|\vec{p}|,p_4)}\,.
\end{align}
Here the denominators are 
\begin{align}
D^2(|\vec{p}|,p_4)\equiv {}& \vec k^2 a^2 A_s^2(|\vec{p}|,p_4) 
+k_4^2 a^2 A_t^2(|\vec{p}|,p_4)\nonumber\\
&+ B_m^2(|\vec{p}|,p_4),\\
\Dcal^2(|\vec{p}|,p_4)&\equiv  \vec k^2 a^2 \A_s^2(|\vec{p}|,p_4)+ 
k_4^2 a^2 \A_t^2(|\vec{p}|,p_4)\nonumber\\
&+ \B_m^2(|\vec{p}|,p_4)\,,
\end{align}
and they still obey
\be
\label{eq:inv1}
D^2(|\vec{p}|,p_4^2) \mathcal{D}^2(|\vec{p}|,p_4^2) =1\,.
\ee
This looks just like the Landau gauge case, but with an extra
structure function. Applying the same line of reasoning as in
Ref.~\cite{Bowman:2001sz, Bowman:2005vx} now leads immediately to 
Eq.~(\ref{eq:stdress}) in the main text.

Turning now to the static propagators, we first note that 
$A_t(|\vec{p}|,p_4)$ is 
an even function of $p_4$, so that the $\gamma_4$ contribution cancels 
by $T$-symmetry when integrating $S^{-1}(p)$ over the (energy) Brillouin zone. 
Therefore: 
\begin{align}
&S^{-1}(\vec p) = i\veckslash a \intpfour A_s(|\vec{p}|,p_4) 
+ \intpfour B_m(|\vec{p}|,p_4)  \label{aa}\\
&= i\veckslash a \intpfour \frac{\A_s(|\vec{p}|,p_4)}{\Dcal^2(|\vec{p}|,p_4)}
+ \intpfour \frac{\B_m(|\vec{p}|,p_4)}{\Dcal^2(|\vec{p}|,p_4)}  \label{bb}
\end{align}
and
\begin{align}
&S(\vec p) = - i\veckslash a \intpfour \A_s(|\vec{p}|,p_4) 
+ \intpfour\B_m(|\vec{p}|,p_4)  \label{cc}\\
&= - i\veckslash a \intpfour\frac{A_s(|\vec{p}|,p_4)}{D^2(|\vec{p}|,p_4)} 
+ \intpfour\frac{B_m(|\vec{p}|,p_4)}{D^2(|\vec{p}|,p_4)}.  \label{dd}
\end{align}
From \Eq{aa} we obtain:
\begin{align}
&S(\vec p) = \frac{-i\veckslash a \intpfour A_s(|\vec{p}|,p_4) 
+ \intpfour B_m(|\vec{p}|,p_4)}{P^2(|\vec p|)},  \label{ee}\\
&P^2(|\vec p|)\equiv  \vec k^2 a^2 
\left(\intpfour A_s(|\vec{p}|,p_4)\right)^2 \nonumber\\
&\hspace*{2cm}+ \left(\intpfour B_m(|\vec{p}|,p_4)\right)^2\,,
\end{align}
while inverting \Eq{cc} yields:
\begin{align}
S^{-1}(\vec p) &= \frac{i\veckslash a \intpfour \A_s(|\vec{p}|,p_4) 
+ \intpfour \B_m(|\vec{p}|,p_4)}{\P^2(|\vec p|)},  \label{ff}\\
&\P^2(|\vec p|)\equiv  \vec k^2 a^2 
\left(\intpfour \A_s(|\vec{p}|,p_4)\right)^2 \nonumber\\
&+ \left(\intpfour \B_m(|\vec{p}|,p_4)\right)^2.
\end{align}
From \Eq{aa} and \Eq{ff} we thus obtain:
\begin{align}
	\intpfour A_s(|\vec{p}|,p_4)&=\frac{\intpfour 
\A_s(|\vec{p}|,p_4)}{\P^2(|\vec p|)}, \\
	\intpfour B_m(|\vec{p}|,p_4)&=\frac{\intpfour 
\B_m(|\vec{p}|,p_4)}{\P^2(|\vec p|)}\,
\end{align}
while \Eq{cc} and \Eq{ee} give:
\begin{align}
	\intpfour \A_s(|\vec{p}|,p_4)&=
\frac{\intpfour A_s(|\vec{p}|,p_4)}{P^2(|\vec p|)}, \\
\intpfour \B_m(|\vec{p}|,p_4)&=\frac{\intpfour 
B_m(|\vec{p}|,p_4)}{P^2(|\vec p|)}.
\end{align}
Writing the static propagator as:
\begin{align}
S(\vec p) &= Z(|\vec p|)\frac{-i\veckslash a + M(|\vec p|)}{\vec k^2 a^2 + 
M^2(|\vec p|)},  \label{gg}\\
S^{-1}(\vec p) &= Z^{-1}(|\vec p|) \left( i\veckslash a + M(|\vec p|)\right).  
\label{hh}
\end{align}
and comparing \Eq{ff} and \Eq{hh} we finally have:
\begin{align}
Z(|\vec p|) &= \frac{\P^2(|\vec p|)}{\intpfour \A_s(|\vec{p}|,p_4)} 
= \frac{1}{\intpfour A_s(|\vec{p}|,p_4)}, \\
	M(|\vec p|) &= \frac{\intpfour \B_m(|\vec{p}|,p_4)}{
\intpfour \A_s(|\vec{p}|,p_4)} =\frac{\intpfour B_m(|\vec{p}|,p_4)}{
\intpfour A_s(|\vec{p}|,p_4)}\,.
\end{align}

%--------------
% Bibliography:
%--------------
\bibliographystyle{apsrev4-1}
\bibliography{references}

\end{document}